%% file: main.tex
\documentclass[conference]{IEEEtran}

\ifCLASSINFOpdf
\else
\fi

\usepackage{tikz}
\usepackage{amsmath}
\usepackage{multirow}
\usepackage{xspace}
\usepackage[hidelinks]{hyperref}  
\usepackage{url}            

\usepackage{breakurl}
\usepackage{booktabs}
\usepackage{subcaption}
\input{definitions}

\begin{document}

\title{
It's TEEtime: A New Architecture Bringing Sovereignty to Smartphones}

\author{\IEEEauthorblockN{ Friederike Groschupp \; Mark Kuhne \; Moritz Schneider \; Ivan Puddu \; Shweta Shinde \; Srdjan Capkun}
\IEEEauthorblockA{ETH Zurich\\firstname.lastname@inf.ethz.ch}
 }

\IEEEoverridecommandlockouts

\maketitle

\input{sections/abstract}
\input{sections/intro}

\input{sections/motivation}

\input{sections/overview}
\input{sections/design}

\input{sections/impl}
\input{sections/eval.tex}

\input{sections/relatedwork}

\input{sections/conclusion.tex}

\bibliographystyle{IEEEtranS}
\bibliography{references}

 \appendices
\input{sections/appx}

\end{document}

%% file: definitions.tex
\newif\ifdraft{}

\drafttrue{}

\draftfalse{}

\ifdraft

\newcommand{\todo}[1]{\textcolor{red}{TODO: #1}}
\newcommand{\shweta}[1]{\textcolor{purple}{#1 -- Shweta}}
\newcommand{\fried}[1]{\textcolor{blue}{#1 -- Friederike}}
\newcommand{\markk}[1]{\textcolor{green}{#1 -- Mark}}
\newcommand{\srdjan}[1]{\textcolor{brown}{#1 -- Srdjan}}
\newcommand{\moritz}[1]{\textcolor{cyan}{#1 -- Moritz}}
\newcommand{\ivan}[1]{\textcolor{orange}{#1 -- Ivan}}

\else

\newcommand{\todo}[1]{}
\newcommand{\shweta}[1]{}
\newcommand{\fried}[1]{}
\newcommand{\markk}[1]{}
\newcommand{\srdjan}[1]{}
\newcommand{\moritz}[1]{}
\newcommand{\ivan}[1]{}

\fi

\newcommand{\loc}{LoC\xspace}

\newcommand{\codename}{TEEtime\xspace}

\renewcommand{\paragraph}[1]{\noindent {\bf #1}}

%% file: sections/abstract.tex
\begin{abstract}
Modern smartphones are complex systems in which control over phone resources is exercised by phone manufacturers, operators, OS vendors, and users. These stakeholders have diverse and often competing interests. Barring some exceptions, users, including application providers, entrust their security and privacy to OS vendors (Android and iOS) and need to accept the constraints they impose. Manufacturers protect their firmware and peripherals from the OS by executing in the highest privilege and leveraging dedicated CPUs and TEEs. OS vendors need to trust the highest privileged code deployed by manufacturers. This division of control over the phone is not ideal for OS vendors and is even more  disadvantageous for the users. Due to it, users are generally limited in what applications they can install on their devices, in the privacy model and trust assumptions of the existing applications, and in the functionalities that applications can have (e.g., restriction over the use of NFC). Effectively, the only mechanisms currently in place to protect users are governmental policies. We show that it is also possible to leverage technical solutions to protect user interests.

In this work, we propose \codename{}, a new smartphone architecture based on trusted execution allowing to balance the control that different stakeholders exert over phones. A more leveled control over the phone means that no stakeholder is more privileged than the others to protect its code and data. In particular, \codename makes users sovereign over their phones: It enables them to install sensitive applications in isolated domains with protected access to selected peripherals (e.g., display) alongside an OS (e.g., Android). \codename achieves this while maintaining compatibility with the existing smartphone ecosystem and without relying on virtualization (and hence without having to trust any hypervisor); it only assumes trust in a phone's firmware. Compared to existing smartphone TEE architectures, \codename is the first TEE architecture that allows isolated execution domains to gain protected and direct access to peripherals, without relying on a hypervisor. \codename is based on Armv8-A and achieves peripheral isolation using a novel mechanism based on memory and General Interrupt Controller (GIC) protection. We demonstrate the feasibility of our design by implementing a prototype of TEEtime, and by running exemplary sensitive applications on top of our prototype. 

\end{abstract}

%% file: sections/intro.tex
\section{Introduction} 
\label{sec:introduction}
Three major stakeholders play a role in the current smartphone ecosystem: manufacturers, OS vendors, and users. They all generally benefit from each other's presence in this ecosystem. 
Manufacturers provide the hardware and firmware that OS vendors rely on to serve users. OS vendors provide a good user experience. Users benefit from the services available on their devices in multiple aspects of their daily lives.
However, these stakeholders tend not to trust each other and thus implement several measures to protect their code and data from other stakeholders.

Phone manufacturers offer several services to the user via their own applications (e.g., camera, file management) and services (e.g., power management, fingerprint authentication). They tend to protect the firmware and peripherals on the phone (e.g., sensors~\cite{appletouchid}, baseband modem~\cite{delugre2011reverse}\footnote{Given that the manufacturers and network operators share the control over the baseband, for this paper, we subsume the control that network operators exert under the label of manufacturers.}) from user apps and OS by executing their code with high privileges. While they could choose to run all their code in firmware, they generally do not, thus avoiding unnecessarily bloating the TCB. Concretely, they follow one of the following three approaches: 1) they run parts of their code on dedicated security CPUs~\cite{applesecureenclave}, 2) they deploy their code in the TrustZone (TZ) secure world~\cite{trusty,optee}, and 3) use VMs to execute trusted apps in isolation from the OS and user apps~\cite{gunyah_qualcomm,hafnium_hypervisor}. Note how in all these three approaches the manufacturer code executes at a higher privilege than the OS, and the phone manufacturer's code can access the user's code and data.

Major operating systems, such as iOS, Android\footnotemark{}, and variants like GrapheneOS~\cite {grapheneos}, have more particular security requirements. They come with a large codebase%
, since they provide a wide range of services to the users, but a large codebase also implies a higher likelihood of vulnerabilities~\cite{androidsecuritybulletin}. Additionally, as discussed above, they have to include the manufacturers code in their TCB. Generally, OS vendors aim to protect at least their security-sensitive services from exploits due to the inherently large TCB.
\footnotetext{For the rest of the paper, Android refers to the Google-controlled variant. While there is the Android Open Source Project (AOSP), it does not come with core features such as Google Mobile Services, which are closed-source and proprietary to Google~\cite{arstechnica_2018,google_gms}.}
To meet their security requirements OSs follow one of two approaches: 1) they rely on Arm TrustZone to execute security-sensitive services~\cite{trusty,optee}, and 2) they employ hypervisor-based solutions, such as pKVM~\cite{pkvm}. Either way, the main goal is to isolate sensitive code and data in a protected TEE environment, which we refer to as \emph{domain}\footnote{The literature uses analogous names such as enclave or realm.}.

In this work, we argue that both approaches are undesirable. TrustZone-based systems have been shown to be susceptible to privilege escalation attacks, where an attacker exploits a trusted app to then compromise Android~\cite{attackingandroidfromtz}. Several attempts to exert fine-grained isolation in the secure world have been deployed, the most notable being secure world partition managers such as the Hafnium hypervisor~\cite{hafnium_hypervisor}. Unfortunately, this approach still suffers from the problem that all code executing in the secure world is overprivileged and can access the memory of the normal world, which includes the normal OS and user applications.

Regarding the second approach, Android recently rolled out pKVM~\cite{pkvm}, a lightweight hypervisor that executes in the normal world. It allows Android to launch security-sensitive apps (e.g., keystore) in its own VM such that any bug in the large Android code-base cannot compromise the VM apps. Since the hypervisor executes with a higher privilege (EL2) than the OS, it is protected from OS bugs while allowing the manufacturer to execute its firmware in the highest privilege (EL3).
However, this prohibits manufacturers from executing their own code in EL2 (manufacturer approach \#3), e.g., type-1 hypervisors~\cite{gunyah_qualcomm}, leading to conflicting requirements between manufacturers and OS vendors.

The directions in which manufacturers and OS vendors are moving ultimately limit the end-users the most. Due to the current security landscape, users can generally install only pre-approved apps (by either the OS vendor or manufacturer). Additionally, apps are restricted in how they can operate on the phone~\cite{larus2021whose}. For instance, OSs limit the API that apps get to access Bluetooth, NFC, or the camera. This limits the range of apps that could be developed to what the OS vendors decide for the users. In terms of their privacy, users depend on and trust the OS and the manufacturer to make the right security decisions by assessing the apps and limiting their access. The OS further provides users with fine-grained access permissions over the apps. However, the user has no option but to trust the OS provider and phone manufacturer with the privacy of its most sensitive data. Overall, the user has to rely on these companies that become arbiters and gatekeepers. Recent examples have shown that OS vendors can exert such control to go beyond security enforcement and lock in users to an ecosystem~\cite{telegraph_2018,theverge_2016,theverge_2020}.

In summary, the current ecosystem is undesirable for all stakeholders. The security solutions that OSs and manufacturers are deploying are neither complete nor compatible with each other (hindering their deployment) and users are limited in what they can do with their devices while still not being in full control of their data and code.

With the aim of achieving a better smartphone ecosystem for all stakeholders, we propose \codename. \codename is a new smartphone architecture that allows to create isolated domains for security-critical apps with full access to selected peripherals. This allows any stakeholder to achieve rich functionality without compromising the security goals of the other stakeholders. In particular, \codename enables user sovereignty over their phones and allows them to allocate compute time, memory, and peripherals to different software domains during the runtime of the device.  Simultaneously, \codename does not allow the users to violate the security of the legacy OSs, nor can the OS subvert the security of the manufacturer code and data. \codename's technical main contribution is that it allows adapting the Arm mobile architecture in a way that peripherals can be directly shared across various domains without having to rely on virtualization. This is important for the smartphone ecosystem as OS vendors and manufacturers (and even users) might not necessarily trust one of the others to control the hypervisor providing the virtualization of the peripherals.

Leveling the playing field by balancing the control that OS vendors, users, and platform manufacturers can exert is a complex technical issue. To realize \codename, we had to overcome four major challenges revolving around resource sharing between mistrusting domains. Several of these challenges emerge due to the unique setting of having to either share or re-assign resources (peripherals, execution time) without utilizing a hypervisor. The first challenge concerns execution and memory isolation. Given the setting in which \codename operates, we cannot have a domain enforce the memory isolation of the others. However, barring denial of service (DoS) attacks, we show that scheduling can still securely be offloaded to other domains. The second challenge is how and when to assign peripherals to domains. In current smartphones, peripherals are pre-assigned to the OS, while in \codename we  re-assign them  dynamically to domains without relying on a hypervisor. The third challenge concerns peripheral isolation: prohibiting mistrusting domains from accessing each other's resources (while still being compatible with Armv8-A). Finally, the fourth challenge we tackle with \codename is how to securely route interrupts to domains on Arm. Arm was designed either assuming a single OS or hypervisor in normal world, thus we had to overcome this limitation to correctly handle interrupts in \codename. To demonstrate the feasibility of \codename and that it is possible to run different domains side-by-side on the same platform without resorting to virtualization, we prototype our design on the Arm FVP simulator. Further, we develop a range of domains with varying peripheral requirements that we run aside a fully functional OS, showcasing the versatility of \codename.

In summary, we make the following contributions: 
\begin{itemize}
    \item We motivate the need for a new smartphone design that balances the interests of smartphone stakeholders (users, manufacturers, and OS vendors) and allows users to be sovereign over their devices. 
    \item We propose \codename, a new Trusted Execution Environment (TEE)-based smartphone architecture as a core component to solving this problem; we argue that \codename can align the interests of users, operators, and OS vendors.
    \item \codename proposes a new peripheral isolation approach based on memory protection and on novel interrupt isolation via the General Interrupt Controller (GIC); \codename is the first TEE architecture that allows isolated domains direct and protected access to peripherals (i.e., without virtualization or intermediary).
    \item We implement a prototype of TEEtime and run several example applications on it: among them secure messaging, biometric authentication, and a VPN service\footnote{A short video demo of our prototype can be found here: https://www.youtube.com/watch?v=m80pTgLjlV8}.
\end{itemize}

%% file: sections/motivation.tex
\begin{figure*}
    \centering

    \begin{subfigure}{.48\textwidth}
        \centering
         \includegraphics[width=0.8\linewidth]{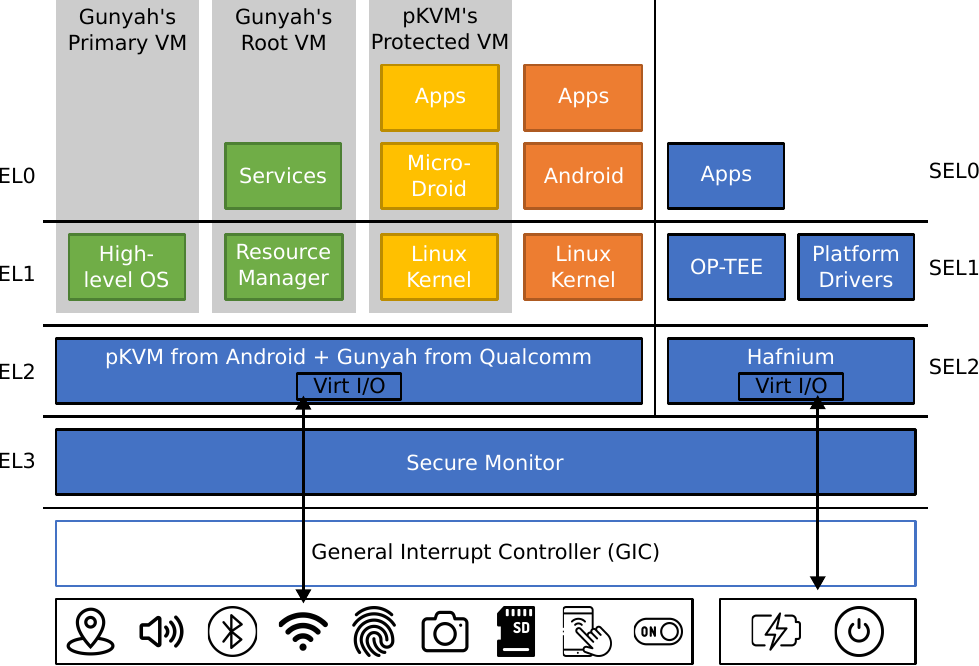}
        \caption{Current ecosystem}
        \label{fig:design_comp_current}
    \end{subfigure}
    \hfill
    \begin{subfigure}{.48\textwidth}
        \centering
         \includegraphics[width=0.6\linewidth]{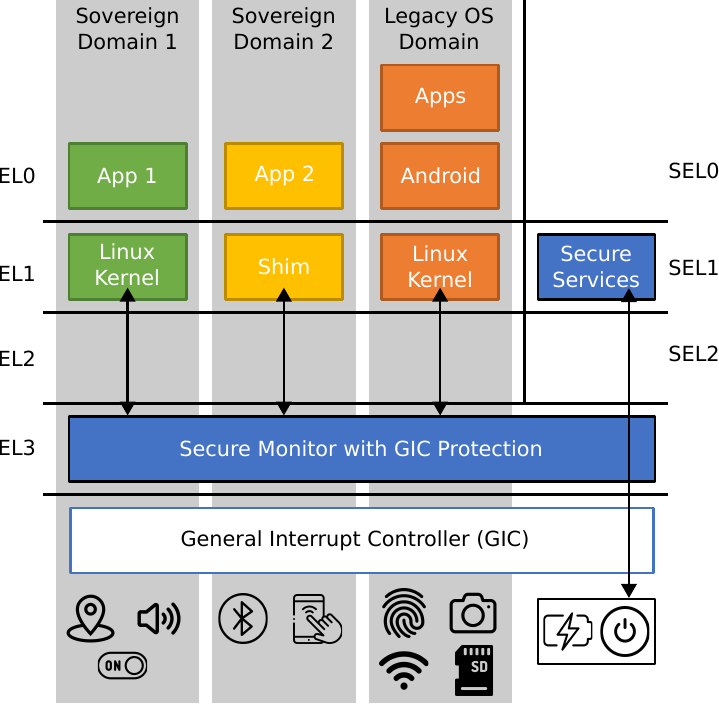}
        \caption{TEEtime}
        \label{fig:design_comp_teetime}
    \end{subfigure}
    \caption{{\bf Design Comparison.} 
    Armv8-A cores offer four privilege layers, from EL0 (lowest privilege) to EL3 (highest privilege). EL3 is populated by the secure monitor (SM). Platforms with TrustZone extensions support secure states SEL0 to SEL2. This processor state can be switched only in EL3. Software running in secure state is more privileged than and protected from software in the non-secure state. (a) Existing designs that add hypervisors, secure OSes, and VMs for isolated execution. (b) \codename creates domains that can achieve the same level of isolation without virtualization. Blue-colored boxes with thick black outlines indicate common TCB. 
    }\label{fig:design_comp}
    \vspace{-1em}
\end{figure*}
\section{Problem Statement} 
In this work, we focus on the problem of balancing control over the smartphone and its resources. Therefore, we discuss examples where the incentives and requirements of different stakeholders, such as OS vendors, phone manufacturers, and users, collide in this section. We highlight the limitations of current phone architectures, motivating the need for a new design. Finally, we generalize these examples into our design goals.

\input{sections/stakeholders}

\subsection{Limitations of Current Solutions}
\label{sec:potential-solns}
The above examples illustrate the tensions between users, OS vendors, and phone manufacturers, which existing phone designs cannot resolve satisfactorily. To summarize, two main aspects currently influence software distribution in the smartphone ecosystem: i) smartphone OSs are generally large monolithic systems (like traditional OSs), ii) sensitive manufacturer code often needs immediate access to peripherals. The consequence of i) is that both OS vendors and manufacturers expect the OS might be compromised eventually; ii) entails that manufacturer code cannot be deployed as a user-level application, e.g., packed as an app store application. 

Together, i) and ii) resulted in architectures in which manufacturers deploy code that is independent and protected from the phone's OS. The most common approach is to utilize the isolation provided by Arm TrustZone  ~\cite{winter2008trusted}. Recently, a second approach is emerging, where isolation is provided by hardware-enabled virtualization instead. We summarize both approaches in Figure~\ref{fig:design_comp_current}.

The first approach is depicted on the right side of the figure. Manufacturers execute applications in SEL0 (secure world EL0) and platform drivers in SEL1 alongside a secure OS (e.g., OP-TEE~\cite{optee}, QSEE~\cite{kai2019guard}, Trusty~\cite{trusty}, Trustonic~\cite{trustonic}, Sierra TEE~\cite{sierratee}, TrustedCore~\cite{busch2020unearthing}). Both run on top of a hypervisor (e.g., Hafnium ~\cite{hafnium_hypervisor}). Alongside the secure world, the main OS (e.g., Android) and its apps execute in the normal world. The secure world is more privileged than the normal world and can access normal world code and data. Note that in this configuration the secure hypervisor isolates only applications executing in the secure world, but does not provide isolation between the secure and normal worlds. Since the manufacturers deploy software running in the secure world, OS vendors are disadvantaged. While they cannot control secure world software, the secure execution of their OS depends on it being bug-free and not malicious.

These concerns are addressed by the second approach, which relies on using hypervisors to provide isolated VMs and virtualize peripherals so that all software can execute in the normal world (left side of Figure~\ref{fig:design_comp_current}). This approach is under active development with Google's pKVM~\cite{pkvm} and Qualcomm's Gunyah~\cite{gunyah_qualcomm}. 
However, the manufacturer's hypervisor (e.g., Gunyah) has to disallow the OS-provisioned hypervisor (e.g., pKVM) to run in the same privilege, thus limiting the security enforcement of the OS. Otherwise, the two hypervisors have to run adjacent to each other without any isolation. This is problematic when OS vendors and manufacturers cannot trust each other. 
Employing a hypervisor in the normal world also means that access to shared peripherals has to be virtualized. Thus, only one software component---either one of the hypervisors or the OSes---has direct access to all peripherals and allows other components to access them via virtual interfaces. Besides the complexity of realizing this at a system level, the solution of using a hypervisor is undesirable from a security perspective. It imposes a design where the virtualization layer can observe all communications with the peripherals and manipulate the peripheral's configurations and run-time behavior (e.g., interrupts). 

Although it has not been adopted yet in any smartphone architecture, a final approach we analyze is Arm Confidential Computing Architecture (CCA)~\cite{arm_cca}. Arm CCA introduces realms: an execution environment that is isolated both from the secure and normal worlds. Adopting the Arm CCA extension would allow manufacturers to deploy software in the secure world and OS vendors to deploy their sensitive apps in a realm with both being isolated from each other. Hence, realms provide the desired security properties to these two stakeholders, but only concerning data and code. Unfortunately, realms are not designed to manage physical interrupts directly and therefore need to trust the OS executing in the normal world and, by extension, the secure world software for peripheral management. Furthermore, all domains need to trust the RMM, a minimal hypervisor managing them.

The friction over the control of the phone's most privileged software layers between manufacturers and OS vendors ultimately leads to the neglect of the end users. They have to accept the conditions that OS vendors and manufacturers dictate. In summary, relying on the secure world or the hypervisors in current smartphones leads to either the manufacturer or the OS vendor having more control over the other's deployed code and data. Hypervisors or virtualizing peripherals in general are not a solution to this problem because only one stakeholder (OS vendor or manufacturer) has control over the virtualization layer, even with Arm CCA. All other stakeholders, including the end users, thus have to trust the virtualization provider.

\subsection{\codename Main Goals} 
\label{sec:goal}

From the discussion above, we derive the need for a new smartphone architecture that balances control in the smartphone ecosystem between the stakeholders. In this work, we propose such an architecture. Our goal is twofold. 

First, we aim to rethink the smartphone architecture to securely distribute the control phone manufacturers, OS vendors, and users have to manage and configure smartphone \emph{resources} such as memory, execution time, and peripheral access. %
A balanced control between these entities means that resource utilization can be \emph{securely} allocated and revoked without the need to rely on other entities. In this context, securely means that the integrity and confidentiality of the data and code of each entity are guaranteed, no matter the configuration currently applied to the smartphone. We believe this goal addresses the shortcomings of existing solutions mentioned in Section~\ref{sec:potential-solns}.

Second, we aim for compatibility with existing smartphone hardware and software stacks. Our solution should work without having to refactor existing operating system code extensively. This benefits all smartphone stakeholders since app stores, OS, and phone manufacturer services can run out of the box, continuing to provide the rich ecosystem that users and developers are used to.

\subsection{Attacker Model}
\label{sec:threatmodel}

In our attacker model, we trust the hardware the phone manufacturer provides, including peripheral hardware and firmware. In particular, we trust the security monitor (SM) and assume its code is public and verifiable. The SM code is either deployed as immutable, or its updates are verifiable by or subject to approval by all stakeholders. 

We assume the attacker can control one or several domains by convincing the stakeholders to install malicious code or by compromising an existing domain. A domain can selectively expose interfaces to provide services to other domains. Domain developers need to decide if they trust the exposed interface of another domain. We make no assumptions about the general trustworthiness of such interfaces or interactions. 

Arm-based phones can be susceptible to side-channel leakage~\cite{zhang2016truspy}. These attacks are out of the scope of our analysis since \codename is not designed to protect against these attacks explicitly. We refer to orthogonal work for mitigations of such attacks~\cite{rane2015raccoon}. Further, an attacker in physical possession of the phone could try to tamper with the execution of domains or attempt to extract their secrets. \codename is not designed to offer any additional protection against these attacks beyond what is available on today's Arm platforms, nor does it introduce any additional vulnerabilities to such attacks. 

We aim to preserve the confidentiality and integrity of the data and execution of domains and their interaction with peripherals. We do not aim to guarantee availability.

%% file: sections/stakeholders.tex
\subsection{Motivation}
\label{sec:stakeholders}

\paragraph{Privacy of User Data.}
Users face uncertainty about when and what kind of data is collected by their phone and how this data is processed~\cite{businessinsider_2019}. While the OS allows users to manage apps' access to peripherals or globally disable access to some resources, such as GPS or Bluetooth, users must trust the OS. Intentional or unintentional misuse of OS-level privileges or opaque policies can endanger user data. For example, Google gathered location information even when the location history feature was seemingly turned off~\cite{apnews_2018, schmidt2018google}.
Ideally, users get design-level privacy guarantees without having to trust the OS.

\paragraph{Closed Ecosystem.}
OS designers have significantly improved the overall security of smartphones~\cite{applesecurity,androidsecurity}. Their efforts include thorough vetting of apps admitted into the app stores, fine-grained permission models, architectural restructuring, the use of secure worlds, virtualization, and secure hardware elements. However, this forces the OS to make a trade-off between the security and  functionality of a phone. Suppose it allows apps to access peripheral APIs freely and does not restrict how they can interact with the system. In that case, this opens users, other applications, and itself to malware attacks. If it implements restrictions, the users and developers will not welcome this decision due to reduced control and functionality. An example is contact tracing apps, where Apple (and to some extent Google) restricted access of government-developed  apps to Bluetooth beacons, citing privacy and performance concerns. This limited the design space and performance of contact tracing apps in several countries~\cite{reuters_2020,theguardian_2020_a,theguardian_2020_b}.

Furthermore, users, developers, and even governments are all subject to app store policies that aim to increase the security and usability of the respective app ecosystem but simultaneously entail restrictions on the functionality of phones. The two largest stores, run by Apple and Google, impose a number of restrictions, including rules on how payment for and within apps must be made and the ban of content deemed unfit at the discretion of the app store provider. Recent examples of such issues are Apple and Google policies which require in-app purchases to use in-store payments, resulting in these companies being accused of gatekeeping in several jurisdictions~\cite{thenewyorktimes_2020}, and OS vendors censoring applications from their stores, either due to self-interest or to state pressures~\cite{bbcnews_2019,reuters_2017}. 
The ability of the users to freely install and run applications without being impeded by the OS would benefit not only the users and developers but would also benefit the OS vendors, which would no longer bear the controversial~\cite{forbes_2021,reuters_2020_b} responsibility of censoring user behavior. Ideally, a phone would offer the possibility to do both: taking advantage of a secure, convenient, and well-maintained ecosystem while running any software without endangering that ecosystem.

\paragraph{Conflicting Privilege Requirements.}
The manufacturer and OS developers usually provide customized firmware for various components/peripherals of the device and trusted applications that deal with sensitive data. The manufacturer ensures that its services (e.g., power management, firmware updates) are safeguarded from all other software on the device. To this end, they typically deploy their services in high-privilege modes. However, this increases the attack surface of user data and the OS.
Analogously, the OS aims to protect its sensitive operations (e.g., processing biometric data, performing proprietary ML predictions based on privacy-sensitive models) from potentially compromised manufacturer-supplied code that executes with higher privileges (since it cannot rely on it being bug-free). 
Ideally, manufacturers and OS vendors can deploy their security-sensitive code, which needs access to peripherals, without increasing  the TCB of the entire system. At the same time, this code must be protected from the rest of the system.

%% file: sections/overview.tex
\section{\codename Overview}
\label{sec:overview}

In this section, we give an overview of our architecture. \codename allows stakeholders to protect their code and data while flexibly sharing smartphone resources by forming isolated domains, which consist of computational resources (similarly to isolated worlds in existing phone architectures~\cite{winter2008trusted}) and, unlike in existing designs, can be assigned protected and direct access to peripherals and their interrupts. \codename achieves this without resorting to virtualization, which is key to reducing trust between various stakeholders. Instead, \codename provides a compact extension to the highest privilege code, the secure monitor (SM), running in EL3 that enforces address space protection and interrupt isolation. The key design goal that guides our considerations is to limit the SM complexity and code size. 
Figure~\ref{fig:design_comp_teetime} illustrates the \codename architecture, which builds on top of Armv8-A. A brief overview over the Arm architecture can be found in Appendix~\ref{appx:background}.

In existing architectures, various software modules execute at different privilege levels and can thus tamper with lower privileged components. With \codename, services installed by users, manufacturers, and OS vendors can run side-by-side in non-secure EL2 to EL0 without the need to trust each other. Through this, \codename allows lifting restrictions of current architectures. In \codename, users can install their software in a separate domain and grant it access to specific peripherals. The software is completely isolated from the OS, thus preventing malware from escalating its privileges. Similarly, the OS and manufacturers can launch sensitive services in \codename domains, thus protecting them from the rest of the software code base. The ability to directly interface with peripherals, without a trusted intermediary, opens up the possibility of moving away from existing non-ideal solutions as we outline in Section~\ref{sec:potential-solns} and Figure~\ref{fig:design_comp_current}. 

In summary, \codename enables users, OSes, and manufacturers to execute apps in a secure and rich environment without having to compromise on security goals and trade-offs (c.f., Section~\ref{sec:stakeholders}). In the following, we provide a brief system overview of \codename with the help of an example deployment. 

\subsection{Illustrative Example: Secure Messaging Domain}
\label{sec:illustrative-example}

Consider a user who wants to run two domains: First, a ``legacy OS'' such as Android to take advantage of its functionality and existing ecosystem. Second, a messaging application that cannot be introspected or interfered with by the legacy OS for secure communication. The messaging domain and its interaction with required peripherals are protected from the legacy OS via address space and interrupt isolation. Given this, the user does not need to trust the legacy OS with their private communication, unlike on today's phones. As this protection goes both ways, the security of the legacy OS does not depend on the benign behavior of the messaging domain. We implemented this and other examples on top of our \codename prototype implementation (cf. Section~\ref{sec:case-study-eval})

To support our messaging domain, the SM grants the messaging domain with exclusive access to the display (trusted I/O) when the user wants to interact with it (e.g., triggered by a button press) and provides it with access to cryptographic material to use untrusted storage (achieved via conventional measurement and sealing mechanisms). The SM will then provide the legacy OS domain with exclusive network and storage access, which the OS domain will proxy for the application domain. This configuration allows the messaging domain to protect the integrity and confidentiality of its user in/output and storage. The communication itself is protected using end-to-end encryption. 

We note that even if this design raises important usability issues, these are beyond the scope of this work. We instead point to the rich literature that shows how one can design hardware buttons and light indicators that allow the users to switch safely between domains~\cite{schechter2007emperor,soriente2007beda}.

\subsection{Platform Configuration}

The target platform of the technical design of \codename is an Armv8A platform with TrustZone extensions. If it is a multi-core platform, we assume that each core has a distinct ID for bus transactions. Further, we assume that the platform is equipped with address space controllers (ASCs), which allow the dynamic configuration of access policies for memory regions and peripherals based on core IDs. For an overview of boards with such controllers, we refer to~\cite{cerdeira2022rezone}. Further, we expect that one interrupt ID belongs at most to one peripheral, which is in line with the configuration of current ARM platforms~\cite{armfvp,imx8m}.
Finally, we assume that EL3 code is either immutable or can be updated only if allowed (i.e., signed) by the manufacturer and/or approved by the user.

\subsection{Isolation of Execution and Scheduling}
\label{sec:exec-resources}
As discussed above, the SM provides isolation of the execution of domains, i.e., it sets up a clean state when domains are launched and saves and restores the domain state when domains are scheduled or preempted (similarly to an OS facilitating context switches). However, the SM does not perform the task of scheduling. Instead, the user delegates this task to one of the domains, which becomes the platform scheduler. This scheduler has the authority to instruct the SM to set up other domains, specifying which resources they are assigned and running them. With this, the SM only enforces the scheduling decisions and is not bloated with complicated scheduling logic itself. Scheduling within a domain (e.g., scheduling different apps inside legacy OS) is still performed by the domains themselves. The platform scheduler is the first domain to run after boot and might also offer an interface to the user to set up new domains.

The isolation of execution can be done in two ways. In \textit{temporal sharing} of the platform, the whole platform executes exactly one domain at a time. The platform scheduler specifies which domain to run and for how long. The SM restores that domain's state, sets up the respective access control mechanisms, starts a secure timer that triggers at the given deadline, and lets the domain execute. Once the timer triggers and control is handed back to the SM, it switches back to the platform scheduler. When multiple cores are available, domains can execute concurrently on separate cores. We call this \emph{spatial sharing}. In spatial sharing, the platform scheduler can dedicate one or multiple cores for running another domain and run multiple domains simultaneously. In this case, the SM provides a clean domain state at the request of the platform scheduler before letting the new domain boot and execute on the assigned cores. In practice, most multi-core systems will support both spatial and temporal sharing of execution. 

Note that using a domain as the platform scheduler has implications on availability guarantees for other domains. The scheduler can refuse to schedule specific or all domains. The user, however, is also in control of this risk and can decide to use another domain as scheduler if they think that it might be trying to censor some or all domains.

\subsection{Memory Isolation}
We use hardware protection based on ASCs to isolate domain memory. Such ASCs generally allow EL3 code to specify access control rules for specific address regions (read/write, non-secure/secure/both/none). Furthermore, if the system is set up accordingly, the controller can define access policies based on the core an access originates from. 

With temporal sharing, only one domain is running at the same time, and we can apply a platform-wide configuration. Therefore, the SM configures the ASC such that all fmemory regions that belong to the currently running domain are accessible (read and write) for non-secure software. When a domain switch takes place, the SM reconfigures the ASC to allow access only to the memory regions belonging to the next domain. When a domain stops execution, \codename re-configures the ASC  such that the domain memory becomes inaccessible again. With spatial sharing, the SM configures the ASC so that each core can only access its domain's memory regions. Note that controllers are able to dynamically partition the memory space, allowing for dynamic creation of and memory assignment to domains during the runtime of the device.

Previous work~\cite{brasser2019sanctuary,cerdeira2022rezone, sun2015trustice} has demonstrated how such isolation can be achieved under temporal and spatial sharing of the platform, discusses technical details and the restrictions and capabilities of different ASCs. \cite{yu2022elasticlave} makes considerations about enclaves sharing memory.  We refer to these works for more details, as memory isolation is a necessary and important component of \codename, but not a novel contribution of our work.

\begin{figure}
    \centering
    \includegraphics[width=\linewidth]{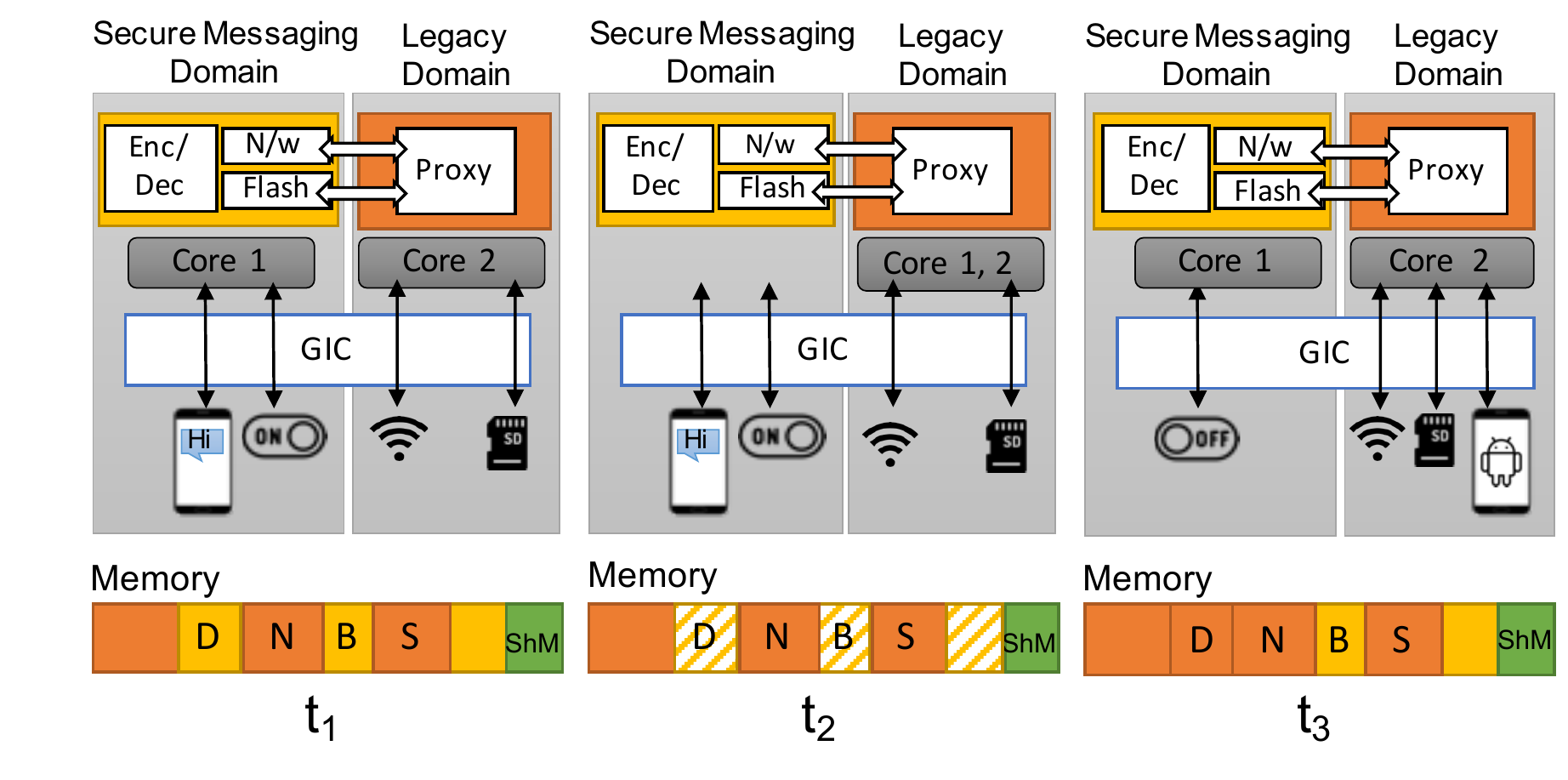}
    \caption{{\bf Example.} Secure messaging domain executing alongside a legacy OS domain. At time $t_1$, display (D) and button (B) in $\tt{ON}$ state are in the message domain executing on core 1. Network (N) and storage (D) are in the legacy OS domain on core 2: the messaging domain can access them via a secure-channel proxy. The memory protection is set such that each domain has its private memory for execution and peripherals, along with a shared memory region (ShM) both of them can access.  At time $t_2$, when the secure messaging domain is scheduled out of core 1, it still owns display and button. At time $t_3$, when the user changes the button to $\tt{OFF}$ state, the display's ownership is transferred to the legacy OS domain. }
    \label{fig:spatial}
    \vspace{-1em}
\end{figure}

\subsection{Peripheral Isolation}
\label{sec:periperals-overview}

\codename allows stakeholders to request and assign the use of peripherals to domains to manage and access peripherals directly without them being virtualized. \codename enforces peripheral isolation using two mechanisms. First, via memory protection, as most peripherals are memory mapped, so their configuration, state, and data can be accessed and updated via memory operations. To achieve this, ASCs are configured to also cover peripheral address ranges to stop domains from accessing an unauthorized peripheral's memory.
Second, through interrupt isolation. Many peripherals are interrupt-driven, meaning a domain needs to be able to configure, receive, and handle interrupts. Since interrupts are managed by a single hardware component (e.g., GIC), this requires that the interrupt system can be shared securely between domains. \codename implements this isolation via GIC protection (see Section~\ref{sec:interrupt_iso_details}). Accessing peripherals via DMA is the second option to interact with peripherals. Protecting DMA access is very similar to standard memory protection. We refer to Strongbox~\cite{deng2022strongbox} for a detailed discussion.

Peripherals are diverse, and while some allow access only by one domain and are rigid in how they are accessed, others have hot-plug capabilities (allowing them to be easily transferred between domains). Even if a particular peripheral instance can only be directly accessed by one domain at a time, the ownership of a peripheral instance can be transferred when needed. Additionally, the domain that owns a peripheral can also act as a proxy for other domains if allowed by the trust assumptions. Furthermore, some phones have multiple instances of the same peripheral with their own memory and IDs, allowing access from several domains at a time. \codename leverages these mechanisms to share peripherals despite the lack of virtualization (see Section~\ref{sec:peripheral_access_modes}).

\subsection{Domain Life Cycle}
A domain consists of a software bundle and a manifest that describes the resource demands of the domain. The domain life cycle in \codename is similar to other TEE-based systems.  On platform boot, the SM runs after a series of trusted boot stages. The SM sets up access control to platform resources and then hands control to the designated scheduling domain.  Domains can be launched at boot (e.g., background services) or as a response to domain requests (e.g., legacy OS initiates a secure authentication domain), or with a user prompt (e.g., start an app). When requested to launch a domain, the scheduling domain chooses a free memory region which becomes the domain's memory, loads the app binary and its manifest into it, frees required resources such as peripherals, and requests the SM to set up the new domain. The SM measures the domain and its configuration and prepares the isolation mechanisms. If the required resources for the domain are unavailable (e.g., because they are assigned to another domain), \codename stops the launch and reports the cause. Once the new domain is set up, the scheduling domain can ask the monitor to run it. The teardown of a domain can only be requested by a domain itself (by invoking the SM) or  by the SM (e.g., triggered by the user), who then frees resources, clears the domain's state and memory, and then resumes the scheduling domain. The memory resources owned by the domain are freed and can be assigned to currently running domains or to new domains.

%% file: sections/design.tex
\section{\codename Peripheral Isolation}\label{sec:arch_description}
The main novelty of \codename is the support for peripheral isolation and sharing without resorting to virtualization. In this section, we provide details of this mechanism. When a peripheral is assigned to a domain, we say that this domain is the owner of the peripheral. Iff a domain owns a peripheral, it is granted two abilities that facilitate the use of the peripheral: (1) Accessing the peripheral and interacting with it and (2) receiving and handling interrupts generated by that peripheral.

Peripherals are usually accessed and configured through memory mapped IO (MMIO) and direct memory access (DMA). To ensure that only the intended domain can access the MMIO of a peripheral, we leverage an address space controller (ASC) as described in Section~\ref{sec:exec-resources}. Such current ASCs for peripherals either work similarly as controllers for memory, where access rights are enforced based on address regions \cite{armtzc400}, or are board-dependent controllers, that work on a per-peripheral basis~\cite{imx8m}. In any case, isolation is enforced similarly to memory: During context switches, the SM withdraws/grants access rights as appropriate. For spatial sharing, these controllers need to be able to enforce access rights based on the accessing core as well. 

While using the ASC solves the issue of isolating access to peripherals, it cannot be used for isolating interrupt delivery and management, as interrupts are configured through one shared interrupt controller and interrupts are routed by the Interrupt Routing System. In existing solutions, the OS and hypervisors are trusted, and hence interrupts are usually delivered to these trusted components~\cite{arm_cca,gunyah_qualcomm,hafnium_hypervisor}, or isolated domains cannot interact with peripherals at all~\cite{brasser2019sanctuary}. In contrast, in \codename, no single domain is trusted to handle the interrupts for all other domains and, given our goals (cf. Section~\ref{sec:goal}), we cannot rely on hypervisors or implement the interrupt management functionality in the SM. Next, we describe the interrupt isolation mechanism we developed for \codename. 

\subsection{Interrupt Isolation Challenges} 
\label{sec:interrupt_iso_details}

Below, we give a brief background on interrupt management in Armv8-A and explain the challenges faced when using the current architecture to isolate interrupts to different domains. Finally, we discuss how we solved these issues while staying compatible with current architectures.

\paragraph{Interrupt handling with Arm GIC.}
The Arm Generic Interrupt Controller (GIC) v3 (and onwards) consists of three main modules: 1) the \emph{distributor} for the configuration of interrupts that are shared among cores, 2) \emph{redistributors} for the configuration of interrupts that are banked per core, and 3) the \emph{CPU interfaces} for the handling of interrupts. The distributor and redistributors are memory-mapped peripherals and collectively known as Interrupt Routing Infrastructure (IRI). The CPU interfaces are accessed through system registers. Identifiers named INTID uniquely define an interrupt and its source (e.g., a peripheral). When an interrupt's source generates an interrupt, its status in the IRI becomes pending. If the INTID is enabled, the IRI's logic forwards the interrupt to a CPU interface based on the state of the system and configuration of the interrupt. The CPU interface signals the interrupt to the core, redirecting its execution to the interrupt handler. Once the software acknowledges the interrupt it becomes active until the end of the interrupt is signaled by the software and the interrupt becomes inactive again. With the default GIC configuration, interrupts are not assigned to any specific core and can therefore be delivered to any core ready to receive interrupts of that priority. However, software can configure the \emph{affinity} of interrupts in the IRI, which designates one specific cores as recipient.

The GICv3 supports the TrustZone extensions and allows interrupts to be configured as secure or non-secure. The inclusion of TrustZone adds two main differences compared to normal interrupt handling. First, non-secure software cannot receive or change the configuration or state of secure interrupts. Second, non-secure interrupts are usually signaled as interrupt requests (IRQs) to the CPU, while secure interrupts are signaled as fast interrupt requests (FIQs), which can only be handled by secure software.

\paragraph{Notion of Interrupt Isolation.}
A domain owns an interrupt iff it owns the peripheral that is the interrupt's source. If a domain is the owner of an interrupt, it requires access to the interrupt configuration and protected routing and handling of the interrupts.
Protected means that the interrupts cannot be observed or modified by other domains, preventing possible information leaks and guaranteeing the integrity of the code executing. The integrity of execution is particularly relevant because we want to be compatible with existing software (e.g., Android), which might have assumptions about the arrival of interrupts. For instance, assume that a domain expects certain interrupts not to be delivered twice in a row (because this cannot happen unless multiple entities can manipulate interrupts). If this assumption is broken, variables might be overwritten with attacker-controlled values. Another problem might arise if a domain suspends operation until an interrupt arrives, e.g., to signal the end of an I/O operation, and another domain suppresses that interrupt.

Therefore, to achieve interrupt isolation, the following guarantees need to hold:
\begin{enumerate}
    \item A domain can configure interrupts iff it owns them.
    \item Only the domain that owns the interrupt can change its state, i.e., set it to pending, active, or retired.
    \item Interrupts only trigger in the owner domain 
    \item Pending interrupts eventually trigger when its owner is executing.
\end{enumerate}

These requirements lead to two main challenges: How can we bind the power to configure an interrupt to the owning domain? And how can we bind the routing and handling of an interrupt to the owning domain?

\paragraph{Challenge I: Binding of Routing.}
Peripherals may trigger interrupts at any time. As described above, when an interrupt triggers, the GICv3 routes non-secure interrupts to any available core; the GIC is unaware of the software context present on each core as our domains are all executing in non-secure state. Usually, this is not a problem if only an OS is present or if a hypervisor is supposed to virtualize interrupts: in both cases, the correct interrupt handlers can be invoked on any core, and the designated handling software is the only one running on the domain. 

With \codename, this is not the case anymore. When the platform is running under temporal sharing, an interrupt may be set pending when the owner is currently not running on the system. The GIC, being unaware of the different domain contexts, would forward the interrupt to the CPU interface anyways. This might cause problems for both domains involved: One domain receives an interrupt that it did not expect and one does not receive an interrupt it did expect. Under spatial sharing, even if the owner is currently residing on a core, the GIC has no knowledge that it needs to route that interrupt to that core. If it picks the wrong core, the same problems as for temporal are caused.
Solving this challenge requires making the GIC domain-aware. We achieve this without making hardware changes to maintain architectural compatibility. 

\paragraph{Challenge II: Domain-Configuration Binding.}
Manipulating the configuration of the GIC suffers from similar issues as the first challenge: the GIC is generally unaware from which domain a request to change interrupt configuration is originating, nor does it know who owns an INTID. Therefore, it cannot enforce access control appropriately. This violates the first two properties of interrupt isolation: ensuring that only the domain that owns an interrupt can set its state (e.g., inactive) and configuration (e.g., disabling interrupts).

While changing the configuration involves MMIO operations, the Arm ASC cannot solve this challenge. Even if the granularity of the ASC would allow enforcing regions with a granularity of GIC registers (i.e. 32 bit), interrupt configuration is tightly packed with up to 32 INTIDs sharing one register. Allowing or blocking access to the GIC as a whole does not work as well, as it would either allow a domain to modify all INTIDs, or prevent it from modifying even its own. To solve this challenge, we need more fine-grained permission enforcement than what is currently available from Armv8-A ASC and GICv3. We propose a solution that does not require hardware changes and only minimal changes in legacy software.

\subsection{Our Approach for Interrupt Isolation}
\label{sec:interrupt_isolation}
We provide two different solutions for supporting interrupt isolation: One which is compatible with temporal sharing and requires no changes to legacy software and one that is compatible with temporal and spatial sharing and requires minimal changes to legacy software. 

\paragraph{Temporal Sharing.}
When all cores are executing a single domain, \codename leverages the fact that a core executing in non-secure mode cannot modify the configurations of secure interrupts. Specifically, we extend the SM with functionality to configure INTIDs of the currently executing domain as non-secure and all other INTIDs as secure. Since the SM performs the domain switches, it can make these interrupt configuration changes before and after domain execution. With this, we solve challenge II.

To solve challenge I, the SM masks INTIDs that are not owned by the currently executing domain. This means that these INTIDs will not trigger while the domain is executing. In particular, domains cannot enable INTIDs of other domains, as the configuration is blocked as described above. This configuration is updated at every context switch. An INTID for arriving for the current domain will then be directly forwarded (if the INTID is configured to be enabled by the domain). An INTID arriving for a suspended domain will be kept pending until the target domain starts executing when it gets unmasked. 

\begin{figure}[t]
    \centering
    \includegraphics[scale=0.54]{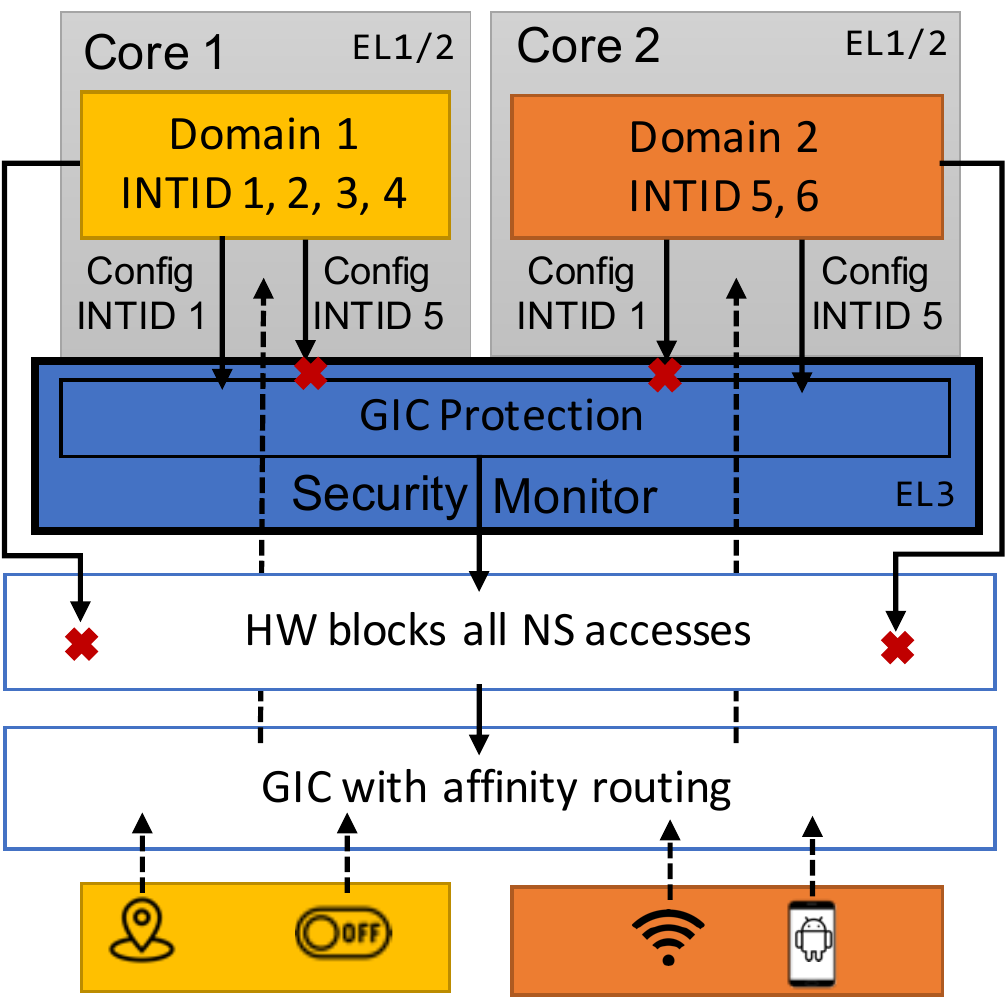}
     \caption{{\bf GIC Protection \& Interrupt Routing for Spatial Sharing.} Domain 1 on core 1 can configure  INTID 1, which it owns, only by calling the monitor. Direct access to the corresponding memory region of the GIC is blocked by the hardware-based access control, as the domain is running in non-secure state. Instead, it must send a request to monitor running in EL3. If the domain requests access to an INTID it does not own, such as INTID 5, the monitor will not perform the access to the GIC.
     When the peripheral triggers an interrupt (shown in dotted arrows), the GIC's affinity configuration ensures that it is routed only to the core that owns the peripheral.
     }
    \label{fig:gic_ac}
    \vspace{-1em}
\end{figure}

\paragraph{Spatial Sharing.}
With cores executing different domains, \codename cannot protect the configuration by marking the interrupts as secure. Doing this would either prevent a domain from legitimately configuring its interrupts or allow other concurrently running domains to break interrupt isolation. Further, the routing behavior of the interrupt would change to that of FIQs because they were marked as secure---which would prevent them from being handled by the domain altogether.
Thus we need a different mechanism to (a) ensure that interrupts belonging to a domain are routed to the core that the domain is executing on (challenge I), and (b) stop domains from changing each other's configurations (challenge II). While this is easier to do with hardware changes to the GIC or the peripheral (see Appendix~\ref{appx:ideal-hw} for one such solution), \codename avoids any such invasive changes. Our solution leverages the SM to do domain-based filtering. 

To protect configuration accesses (challenge II), \codename blocks \emph{all} non-privileged accesses from non-secure software to the GIC using the Arm ASC. Access is still allowed from the EL3 where the SM executes. Hence, to configure interrupts, instead of accessing the GIC directly, a domain needs to perform a secure monitor call (SMC). 
This design choice will require changes to the domain software to invoke monitor calls instead of simply accessing the MMIO. Since our goal is to avoid such invasive changes, \codename implements a transparent mechanism to hook all GIC updates in the software and perform the call automatically (cf. Section~\ref{sec:gatekeeper-module}).
On receiving such a call from the domain, the \codename monitor combines information about which core the request comes from, which domain is running on that core, and INTID ownership to decide whether access is allowed. If it is allowed, the monitor performs the configuration change on behalf of the requesting domain. Figure~\ref{fig:gic_ac} depicts the overview of our access-control placements with an example. While this mechanism can be expensive if the domain performs frequent interrupt configuration changes, we note that they are mostly performed while a device is booting. At runtime, interrupts are not frequently reconfigured, except in cases when a peripheral is hot-plugged or load is being adapted for system-wide load balancing.

To solve challenge I, we observe that the Arm GIC supports the concept of affinity, allowing normal-world software to specify to which target core an interrupt should be routed. Traditionally, this is used to balance the interrupt load of a system, increase overall system performance, or provide low latency for certain tasks, e.g. network handling. \codename re-purposes this feature to specify that an interrupt should be routed to a specific core. Crucially, only the monitor can change INTID configuration and thereby its affinity. It performs assigns the correct interrupts to the correct core every time a domain is executed or suspended. 

\subsection{Peripheral Access Modes}\label{sec:peripheral_access_modes}
With the mechanisms described above, a peripheral and its corresponding INTIDs can be owned by one domain at a time. This might restrict the applicability of \codename to many use cases  e.g., having multiple domains with access to the network or different domains being user-facing after each other, i.e., controlling the display. Therefore, \codename considers and supports different peripheral access modes, whose applicability relies on domain needs and peripheral types. Which modes a domain supports for peripherals can be included in the domain manifest. Note that not all modes are sensible to use and provide adequate security for all peripherals and use cases. We describe the requirements, security, and developer effort  trade-offs below, but leave their standardization for future work. In the end, we envision developers stating the requirements of a domain and the user deciding to allow that during installation based on their trust assumptions (as is done with permissions in app stores today). Below, we give examples of such considerations. For our prototype, we demonstrate how different peripherals are used in appropriate modes (cf. Table~\ref{tab:poc_peripherals}).

\paragraph{Exclusive Access.}
A domain can request a peripheral exclusively (e.g., GPS, flash storage), such that only this domain can use the peripheral during the platform's runtime. The peripheral needs to be available at domain setup.  While every peripheral type supports exclusive access and it does not require tailored functionality in software, this mode is also the most restrictive one. However, reasoning about the security of the peripheral usage is straight-forward, as by design no other domain can access the peripheral at any point and interfere.

\paragraph{Multiplexing.}
Phones can have multiple instances of the same peripheral (e.g., UART ports, buttons, LEDs) with each of them having their own memory and  INTIDs. Similarly, peripherals may natively support multiplexing in an isolated manner (e.g., timers, counters), i.e., have different logical instances with separate memory regions and interrupt IDs. \codename can exclusively assign individual instances to different domains, while it appears to the user that multiple domains have concurrent access to the same peripheral. As long as the different instances of the peripheral are properly isolated, the same security considerations as for exclusive access hold. However, this mode depends on the platform configuration.

\paragraph{Handover.}
Some peripherals (e.g., display) support hot plug, which allows a software-based reset without requiring a system reboot. Ownership of such peripherals can be transferred between domains during the lifetime of domains. A transfer of the peripheral can happen with user interaction (e.g., pressing a button to transfer the display) or by a domain voluntarily ceasing ownership when no longer requiring the peripheral. Once available, another domain that is allowed to access the peripheral can be granted its ownership.

For handover, a few security issues arise. First, if the peripheral contains sensitive state (e.g., the current state of a frame buffer), this state needs to be cleared to make sure that information is not leaked across domains. To reduce the complexity of the SM, this cleanup is delegated to the owner domains. Second, the user needs to be aware of which domain currently owns which peripheral. Such information can, e.g., be communicated through LEDs that can only be controlled by the monitor.

\paragraph{Read-only Access.}
Some devices only provide information and optionally generate interrupts when the information is updated (e.g., GPS, gyroscope, accelerometer, magnetometer). They require minimal and often one-time configuration. In such cases, \codename can allow one domain to perform configurations and receive interrupts. Additionally, it can selectively allow other domains to read the memory where the device writes updates. This allows several domains to have shared access to the peripheral without corrupting the state of the peripheral. 

This approach requires might require adapting the behavior of the read-access domains, as they will not receive interrupts from the device and need to poll in intervals instead. Furthermore, these domains have to trust the owner domain with the configuration of the peripheral---while it might be possible to verify the configuration, they cannot intervene in case of misconfigurations.

\paragraph{Proxy Access.}
The owner domain can decide to allow other domains to have proxy-based access to its peripherals (e.g., storage). In this mode, \codename is not actively involved in enforcement. Instead, the owner offers an API to other domains via inter-domain communication based on shared memory. While this mode does not put any requirements on the peripheral, it requires software changes for all domains---in the owner domain to offer a peripheral-specific API and in the requesting domains to use the domain as proxy instead of directly interacting with a peripheral. To implement the API, different approaches can be taken, ranging from the owner `virtualizing' the peripheral by exposing the same interface the peripheral would, to offering a high-level API that abstracts the peripheral's major operations. 

In proxy mode, domains have to trust the owner domain to manage the peripheral, which means correctly forwarding its functionality and not withholding access. In particular, domains need to be able to attest that a domain is the legit owner of a peripheral with the help of the SM and the owner might restrict access to its API to certain domains. This can be achieved by mutual attestation facilitated by the SM and the measurement made at domain setup. If supported, they can use encryption and integrity protection to exchange data with the peripheral or remote endpoints over a secure channel (e.g., storage, network). At the same time, the owner needs in- and output sanitization techniques to protect itself, the peripheral, and other domains from a misbehaving domain. The interrupts are still routed only to the owner domain; the owner can decide if they want to expose the interrupt configuration and/or relay interrupt information.

\subsection{Peripheral Access Mode Examples} \label{sec:periph_access_modes}
We discuss display and network to illustrate how they fit within the sharing modes explained above. Other peripherals can be similarly adapted.

\paragraph{Display.}
In modern smartphones, the display is the I/O for the user and the main point of interaction with software. Therefore, nearly every domain that requires user interaction will require access to the display at some point. In a usual setup, we expect the display to be used in handover mode, where the user can decide to assign the display to domains through the press of a button. Other options include proxying if the owner is trusted with the confidentiality and integrity of other domains' display data, or multiplexing if the display hardware supports assembling one frame buffer for isolated software. An exclusive mode would only make sense when the user intends to only run domains that resemble background processes. 

\paragraph{Network.}
Network peripherals such as WiFi, cellular, or Bluetooth, are an example for peripherals where it is usually possible to establish a secure  channel (e.g., TLS) with an endpoint, e.g., a remote server. This makes this peripheral a prime candidate for the proxying mode: The owner domain forwards traffic between the domain and an entity outside the phone, thereby acting as just another untrusted hop on the routing path. The domain does not need to be trusted for confidentiality and integrity. As mentioned above, the trade-off compared to using another peripheral access mode is the non-network-owning domains lose availability guarantees. Other reasonable options for network include exclusive access when only one domain requires access to a specific network type or handover in cases where the user wants a domain to access the network for specific purposes and times. One such use case could be assigning Bluetooth generally to a legacy OS domain to listen to music. When the user wants to use a separate domain for Bluetooth-based passkey authentication~\cite{fido_spec,google_passkeys}, ownership is handed over to this domain for the duration of the authentication process.

%% file: sections/impl.tex
\section{Implementation}
\label{sec:impl}

We prototype \codename on Arm Fixed Virtual Platform (FVP) that emulates Armv8-A, a GICv3~\cite{arm_gicv3}, and a TZC-400 Address Space Controller (TZASC)~\cite{armtzc400}, with 8 cores. The FVP configuration does not meet our platform requirements in two aspects. First, the CPU cluster shares one ID among cores, allowing us to enforce memory isolation for temporal sharing but preventing us from enforcing core-based access control for spatial sharing. We refer to Sanctuary~\cite{brasser2019sanctuary} for a demonstration of how this can be achieved with Arm Fast Models. Second, even though it is also designed to arbitrate peripheral accesses~\cite{armtzc400}, the TZASC is not on the access path of the peripheral bus. Therefore, we cannot enforce access control to peripherals. Nevertheless, our SM implementation populates and stores the needed configuration structures to emulate the required functionality of the SM\footnote{Further, we verify that the domains correct execution does not rely on accidentally accessing peripherals it does not own by removing the respective peripheral from the device tree for peripheral and proxy mode or by (un)loading drivers for handover mode.}. Development boards usually come with ASCs for peripheral protection~\cite{imx8m,xilinx_ultrascale}. However, while it is feasible to execute \codename on a hardware board with Arm cores, we were not able to find a board that satisfies all our requirements as detailed in Appendix~\ref{appx:boards}. When porting our implementation to a platform that satisfies all our requirements, we expect the only remaining steps to be (i) adapting the configuration format to the ASC present and (ii) storing the configuration to the actual ASC.

\subsection{Secure Monitor}
\label{sec:domain-dispatcher-module}
To facilitate domain setup, isolation, and teardown, we extend the reference implementation for EL3, the Arm Trusted Firmware-A (TF-A v.2.4). Below, we explain how our extensions facilitate a domain's lifecycle. We use SM to refer to the \codename-extended TF-A.

In our implementation, \codename launches a full Linux kernel (v.5.10.107) as the first domain after boot, makes it the scheduler, and gives it access to all platform resources. The user can then request the setup of a new domain from within this domain. For this, it expects a pointer to a domain bundle consisting of a binary and its peripheral requirements. In our setup, these bundles are deployed into the file system of the kernel during build time, but they could also be downloaded or otherwise supplied during the device's runtime. The kernel then loads the binary into a memory region and prepares the required resources by freeing them. To launch the domain, the kernel invokes the \codename SM API and details the domain's resource assignment (i.e., memory region, core to run on, assigned peripherals and interrupt IDs, initial register state). The SM then measures the binary and manifest, sets up isolation mechanisms, and jumps to the entry point to execute the newly created domain. Peripheral ownership is implicitly stored in the SM by the isolation configuration. In the case of temporal sharing, the SM further sets up a secure timer that will trigger and preempt the domain after a period specified by the platform scheduler.

When a domain wants to pause execution or is preempted, the SM facilitates the domain switch. This can happen by the domain invoking the SM or by the deadline set by the platform scheduler expiring. Besides simple save restore of register state on switches, the SM tracks each domain's access permissions to memory, peripherals, and interrupts via dedicated data structures and configures the TZASC and GIC accordingly. When a domain is rescheduled and resumes execution, the monitor restores its state, including interrupt configuration, which triggers all enabled and pending interrupts in order of decreasing priority once the domain resumes. \codename tears down a domain when the domain or a user requests it. It is the responsibility of the domain to clear residual sensitive information from all resources. Then, the SM configures the domain's resources as free.

\subsection{GIC Protection and Backwards Compatibility}
\label{sec:gatekeeper-module}

Existing OSes and services assume direct access to GIC and configure it via reads and writes. With \codename, this is the case under temporal sharing as well. Under spatial sharing, however, \codename's GIC protection disallows such direct access. Instead, it offers an explicit GIC configuration interface in our secure monitor, which can be invoked with secure monitor calls (SMC). The SMC expects the address and access type as parameters; for write accesses, it additionally expects the value to be written. Using such an interface instead of one with semantic information (e.g., set INTID 37 to \textit{enabled}), allows for straightforward processing in the SM as it mirrors how software currently accesses the GIC. The SM infers the requesting domain and its access permissions from its internal data structures to track domains. 

Based on the permissions and the type of GIC register being accessed, SM assembles a bit mask, in which bits allowed to be accessed are set to one. This is necessary because registers are shared between interrupts, e.g., the interrupt-enabled registers are 32-bit and hold the state of 32 interrupts. For write accesses, this mask and the old value of the register are then used to compute the new value. For read accesses, the mask is used to set inaccessible bits to zero. Ignoring writes and returning zero is on par with specified GIC behavior for unauthorized accesses.

We address three important details in our implementation. First, the SM is operating in secure state and has to perform secure accesses to the GIC. Otherwise, the access would be blocked by the ASC. This requires careful evaluation of some registers where the GIC presents different views to secure and non-secure accesses. Furthermore, we must ensure that domains cannot allocate INTIDs to each other, maliciously or accidentally, as this would violate our requirements for interrupt isolation. Therefore, the monitor ignores requests to route interrupts outside the domain, e.g., when a domain wants to turn off affinity routing or set the affinity of a core it does not own. Third, the monitor must synchronize write access across cores, e.g., by using locks, to avoid race conditions. 

\paragraph{Compatibility.}
Legacy software might not be aware of the GIC protection and will try to access the GIC directly. In \codename, this causes an asynchronous abort---due an illegal non-secure access to an address---that goes to the monitor. In the abort handler, the monitor could check if the access is allowed and, in that case, perform it on behalf of the domain. Unfortunately, the information provided by an asynchronous abort is insufficient to reconstruct the attempted memory access. More importantly, existing code may not block after the GIC access; it proceeds with execution immediately after dispatching a write transaction on the interconnect without waiting for a confirmation. As a result, the abort arrives with few instructions afterward. This makes it challenging to reconstruct the faulting instruction. Even if we can infer the instruction and the register containing the value to be written, the state of this register might have changed between those few instructions. Reconstructing the attempted write is thus unreliable. 

While one can retrofit explicit monitor calls into the existing code, doing this for large code bases is labor-intensive and prone to errors. Instead, \codename provides a minimal layer in EL2 placed between legacy software and the monitor that domains can deploy at their discretion. It then uses \textit{synchronous aborts} to reconstruct memory accesses. In contrast to asynchronous aborts, synchronous aborts trigger immediately and in the faulting core and provide precise information about the fault. To trigger the fault, the EL2 layer sets up the stage~2 translation tables to deny access to the GIC memory regions but leaves the configuration of the rest of the address space untouched. In addition, it comes with a minimal exception handler that parses the synchronous abort syndrome register and calls the monitor. It is important to note that this EL2 module provides compatibility for legacy EL1 software and is part of a domain. It does not perform a security function and is not in the system TCB: the hardware still enforces that software in non-secure state cannot access the GIC.

\subsection{Legacy Software as Scheduling Domain}
\label{sec:retrofit} 
By default, \codename designates a full Linux kernel as the scheduling domain to reuse its existing functionality.
We implement a loadable module to extend the Linux kernel with the capability to launch and schedule domains. 
Every domain is shipped with an accompanying user-level app that executes in the scheduling domain and invokes the module to launch and run the domain. In this way, the app acts as a stand-in for the domain for scheduling---when the kernel scheduler schedules the app, it implicitly schedules the domain. If the domain yields, it will only be resumed when the user app requests it. If the domain is preempted, the user app invokes the kernel module again to request rescheduling. \codename performs domain switches only from the scheduling domain to another domain and back.

%% file: sections/eval.tex
\begin{table}
\centering
\caption{{\bf TCB Impact.} TEEtime in lines of code compared to pKVM\cite{pkvm}, Hafnium\cite{hafnium_hypervisor}, OP-TEE\cite{optee}, and Qualcomm Gunyah\cite{gunyah_qualcomm}. Estimate based on the latest respective release and a best-effort removal of unused code, which is especially challenging for pKVm bundled with the Linux kernel.
}\label{tab:loc}

\begin{tabular}{rrrrrr}
\toprule
 
pKVM  & Trusty & Hafnium & Gunyah & OP-TEE & \bf{\codename} \\
\midrule
10,411 & 18,738  & 22,974   & 32,610  & 73,913  & \bf{859} \\
\bottomrule
\end{tabular}
\vspace{-1em}
\end{table}

\section{Evaluation}

We implemented various domains with different peripheral requirements to  demonstrate that \codename can realize a diverse set of peripheral-based stakeholder use cases, evaluate the effect of \codename on the lifecycle of domains, and estimate the impact of \codename when executing different workloads and benchmarks that operate on peripherals.

\paragraph{Setup \& Methodology.}
We used the Armv-A Base RevC AEM FVP~\cite{armfvp} to simulate 8 Arm cores with 4\,GB RAM, 3 UARTs, 8 LEDs, 8 buttons, Ethernet, flash storage, display, and mouse/keyboard as input.
\codename's secure monitor uses the hardware to generate a domain-specific root key based on its measurement and a device key stored in secure non-volatile memory. Domains can retrieve this key during their runtime.
We use a software-based AES implementation for encrypting and decrypting data at the domain boundary.
For our case studies, we report the setup and tear-down times, along with the detailed breakdown of \codename operations such as GIC configuration and domain switching costs. The emulator provides a detailed cycle-level breakdown of all operations that \codename performs. We use these measurements to calculate relative overheads. As the FVP model is not cycle-accurate, these measurements might not reflect actual overheads when executed on hardware. However, we believe there is still value in collecting this data to understand the general impact of \codename on a system.

\paragraph{TCB.}
The existing TF-A code comprises 54625 \loc. We add 859 \loc to the TF-A to realize the \codename SM described in Section~\ref{sec:domain-dispatcher-module}. 
Since all existing approaches need to trust the TF-A in addition to other software they deploy (e.g., hypervisors), \codename SM only minimally increases the TCB, as summarized in Table~\ref{tab:loc}.

\paragraph{Non-TCB.}
\codename adds  untrusted code outside the TCB, as summarized in Sections~\ref{sec:gatekeeper-module} and \ref{sec:retrofit}. We add 855 \loc to interface with the GIC, specifically in u-boot to set up the EL2 runtime stage2 translations and new exception handlers. \codename adds a kernel module to the scheduling domain to launch spatial and temporal domains (157 \loc).

\subsection{Case-studies} 
\label{sec:case-study-eval}

We demonstrate the versatility of \codename with six domains written from scratch or ported from existing code bases, in addition to the legacy OS domain. Our domains consist of apps running as user applications on top of a ramdisk-based Linux kernel that acts as a runtime.
Table \ref{tab:poc_peripherals} shows the app LoCs and binary sizes, along with a summary of peripherals, modes, and number of distinct INTIDs used in each case study. Appendix~\ref{appx:screenshots} shows screenshots of our apps running on \codename, a short video demo can be found here\footnote{\url{https://www.youtube.com/watch?v=m80pTgLjlV8}}.

\paragraph{Secure Data Vault.}
The purpose of the app is to persist data such as secrets (e.g., private keys), app-specific sensitive information (e.g., credit card information~\cite{samsung-pay}), and user data (e.g., chat history) while flash storage is owned by another domain. We implement the legacy OS to act as disk proxy to persist data to and retrieve from storage. The domains communicate via a simple API that supports reading and writing data blobs identified by filenames to storage. If the data requires confidentiality, the vault domain encrypts the data before sending it to the disk proxy. We reuse this app as building block for other domains.

\paragraph{Biometric Authentication.}
Fingerprint authentication requires access to the fingerprint sensor as well as reading, storing, and checking biometric information. While OS and apps rely on  biometric authentication, they prefer not to be responsible for securing it to avoid legal repercussions~\cite{gdpr}. Therefore, the task is outsourced to a dedicated domain, such that neither the data nor the comparison computation can be compromised by the OS or rogue apps. Our biometric authentication domain prompts the user to register their fingerprint. Since no fingerprint sensor is available on the FVP, we create a random bitstring and treat that as user input. We save it using the data vault described above. After setup, the app runs in the background and waits for authentication requests from the legacy OS. When invoked, it reads the fingerprint from a sensor (emulated by bit strings read from a memory location) and performs a fuzzy string match to compare it against the user's fingerprint retrieved from the vault. It then informs the legacy app if the authentication was successful. While phones have dedicated secure elements that achieve such authentication, our app demonstrates that \codename can serve as a replacement or augment the security of existing authentication mechanisms.

\paragraph{System Configuration \& Status.}
This app is a service from the manufacturer to report the state of the system to the user and sync with external services to fetch updated information about updates, stolen devices, and leaked keys. Our app connects to a pre-configured trusted server to query  whether any firmware updates are available. It has exclusive access to the network to perform this task and requests the OS to hand over the display to show the firmware information to the user securely. 
\codename changes the color of the LED to assure the user that the display is indeed controlled by the app's domain.

\begin{table}[t]
\centering
\caption{{\bf Summary of Case-studies.} For the six domains, Column 2 shows the peripherals and their access mode. Columns 3 and 4 show the lines of code and the compiled binary size of the apps running within the domain. Column 5 depicts the number of INTIDs the domains are assigned ownership of at some point during their liftetime.
We represent the modes as E: Exclusive, H: Handover, M: Multiplexing, P$_L$: Proxy in legacy domain, P$_A$: Proxy in app domain for storage (Stor), UART, network (NW), display (Disp), button (Btn), and mouse and keyboard (M\&K).
}
\resizebox{0.45\textwidth}{!}{%

\begin{tabular}{@{}llrrrrr@{}}
\toprule
Case-study      & Peripherals Modes                 & LoC      & Size                 & INTID\#  \\ \midrule
Vault           & Stor (P$_L$), UART (M)            & 873      & 4.891\,MB            & 1       \\
Auth            & Stor (P$_L$)                      & 1,436    & 7.996\,MB            & 0       \\
Firmware        & NW (E), Disp (H), Btn (H)         & 209          & 0.750\,MB            & 1       \\
Browser         & NW (P$_L$), Disp (H), M\&K (H)    & 709,921     & 4.056\,MB            & 2       \\
VPN             & NW (P$_A$), UART (M)              & 2,142      & 0.034\,MB            & 2        \\
Messenger       & Stor (P$_L$), NW (P$_L$),         & 637,036       & 3.882\,MB            & 2   \\
                & Disp (H), M\&K (H)                &            &                      &          \\ 
\bottomrule
\end{tabular}%
\label{tab:poc_peripherals}
}
\vspace{-1em}
\end{table}

\paragraph{Secure VPN.} 
VPNs operate by modifying the headers of IP/TCP packets, e.g., by segmenting TCP packets, injecting SYN packets, or adapting sequence numbers~\cite {wang2020symtcp, bock2019geneva}. Usually, the OS has to implement this functionality network stack and then allow VPN apps to access it. Instead, we implement a VPN app that has exclusive access to the network peripherals. Any domain that wants to use the network can explicitly connect to the VPN domain to send and receive packets. We demonstrate that our VPN app can transparently tunnel the legacy domain's network connection, thus requiring no changes to the existing kernel and apps. We port an existing application (ProxyLite~\cite{proxy-lite}) and reconfigure it to execute in a sender-receiver setup to serve as a VPN service between the legacy and the application domain.

\paragraph{Secure browsing.} 
With this app, the user can browse the web and upload or download content in a way that the legacy OS cannot observe or tamper with the communication even though it controls the network. We port an existing GUI-based web browser, Dillo~\cite{dillo}, for the app domain to access the Internet using the legacy domain as a network proxy. A user interacts with the browser via the display, mouse, and keyboard which are handed over from the OS. The program is rendered using the window manager Fluxbox~\cite{fluxbox} executing on top of the Xorg~\cite{xorg} display server in the app domain. Since we terminate the connections inside the domain, the app domain maintains the session keys and performs the en- and decryption of traffic.

\paragraph{Secure messaging.} 
We implement the illustrative example described in Section~\ref{sec:illustrative-example}. Our app connects and exchanges messages with a local server, via a network connection proxied by the legacy OS domain. Messages between the server and the app are encrypted. Lastly, we hand over the display, mouse, and keyboard from the legacy domain to the app domain. The app is GUI-based and built with the FLTK~\cite{fltk} library. Similar to the web browser it is rendered using Fluxbox on top of Xorg in the app domain.

\subsection{Lifecycle Breakdown}

We measure the lifecycle overheads of \codename with two custom workloads. We start by implementing a simple app that executes bare metal. %
The app has exclusive access to UART via a small driver included in the domain. It takes user input from UART (receiver channel) and responds by echoing it back (sender channel). Next, we boot an unmodified kernel in a \codename domain next to the legacy OS. 
We configure the kernel with two representative runtime environments bootstrapped using an initramFS: A minimal example with a fully custom initramFS using BusyBox~\cite{busybox} as the init process and a more extensive initramFS environment created using Buildroot~\cite{buildroot}.

We observe that load and setup costs depend on the size of the domain binary, while domain setup costs depend on the app-intrinsic complexity. Temporal execution takes longer in general because of the additional context switches that are incurred every time the domain is scheduled in and out. The GIC protection employed by \codename incurs overheads when the domain attempts to access the GIC. For our kernel, we observe a total of 37 reads and 19 writes when we boot the unmodified kernel as the legacy OS. We see 25 reads and 329 writes when we boot the kernel in the domain next to the legacy OS. The number of writes increases when booting as a domain because it attempts to re-write the values blocked by configuration memory. After boot, we do not see any reads or writes to the GIC configuration memory, as the handling of interrupts is done through system registers as expected.
Detailed results can be found in Appendix~\ref{appx:lifecycle}.

\subsection{Performance}

Domains observe different performance overheads depending on their peripheral access modes. We systematically measure this impact on three common types of workloads and summarize the main results below.
Appendix~\ref{appx:perf} reports the detailed performance breakdown.

\noindent \textbf{Storage.}
We set up a benchmark that performs reads and writes of different sizes varying from 1\,byte to 1\,MB files. 
For the baseline, we run the benchmark in a domain that has exclusive access to the storage.  
We then execute the benchmark in a domain that accesses storage through a proxy.
When compared to the baseline that accesses the storage in plaintext, we observe an average  slowdown of 8.60$\times$ and 3.54$\times$ for reads in temporal and spatial modes, 7.96$\times$ and 3.31$\times$ for writes. 
We attribute this mainly to memory copy operations; temporal execution is slower than spatial because of the high number of context switches. When we turn on encryption and domains, we observe an average slowdown of 2.18$\times$ for reads and writes in spatial mode, and 4.39$\times$ for reads and writes in temporal modes. 

\noindent \textbf{Network.} We test the network performance with Apache Bench~\cite{apache-bench} by accessing a standard Apache server landing page on the host system of the emulator; we perform up to 30 concurrent connections capped at 900 total connections.
We execute the benchmark in baseline, exclusive, and proxy access modes. 
When domains access the network in proxy mode, we observe a 10$\times$ slowdown. Most of this overhead is due to socket handling in our ProxyLite port.

\noindent \textbf{System.}
We evaluate the impact of \codename on the system efficiency of the legacy OS domain. 
We execute an OS stress test benchmark, LMBench~\cite{lmbench}, in the legacy domain under three configurations.
A baseline where the OS is the only domain on the platform, one where the OS domain is executing alongside an app domain that is executing an intensive AES computation, and one where we run the benchmarks in the app domain. We configure the legacy domain to have access to all peripherals, while the app domain executes on another core in spatial mode and only uses CPU and memory. We observe no noticeable change for the legacy domain (0.08\% bandwidth gain) and a 9.4\% bandwidth loss for the application domain.

%% file: sections/relatedwork.tex
\section{Related Work}
\vspace{-0.5em}
Operating system vendors and manufacturers of mobile phones are utilizing two main approaches, namely virtualization and secure world applications (as shown in Figure~\ref{fig:design_comp}). While operating system vendors promote the use of virtualization to minimize the trusted computing base (TCB) in the secure world, manufacturers prefer TEE-based approaches as they do not want to add a hypervisor to their TCB. The user has little leverage in the current smartphone architecture. Nevertheless, these two approaches have notable drawbacks. While virtualization properly supports peripheral access, it replicates the issues currently existing with the OS but in a different layer, and thus, uses flawed trust assumptions. TEE architectures, on the other hand, provide stronger trust assumptions but usually over-privilege the TEE/secure world, and they usually lack support for peripheral access for enclaves.

Virtualization-based approaches rely on a hypervisor in EL2 that virtualizes the underlying hardware resources for virtual machines running in EL1. Google's pKVM~\cite{pkvm}, Hafnium~\cite{hafnium_hypervisor}, and Qualcomm's Gunyah~\cite{gunyah_qualcomm} are some of the hypervisors that exclusively target the Arm architecture. They all contain some memory management; some contain a scheduler or drivers to virtualize peripherals. All three of these are designed to be small (see Table~\ref{tab:loc}). pKVM and Hafnium delegate the scheduling and virtualization of peripherals to the host virtual machine, whereas Gunyah includes its own scheduler. Virtualizing peripherals is delegated to the host OS in all three hypervisors. In all of these, the host OS cannot inspect the memory of other virtual machines, but it observes and may manipulate any communication with peripherals.
Arm CCA~\cite{arm_cca} is a recent Arm architectural feature that allows launching virtual machines in a separate realm mode. It was designed to enable more secure virtual machines, as anything in the realm mode is not accessible to any other mode. However, scheduling and peripheral access is still delegated to the host OS, thus the same limitations of virtualization-based approaches apply.

While there exist many TEEs designed for mobile phones, we first analyze two open-source industrial proposals: OP-TEE~\cite{optee} and Google's Trusty~\cite{trusty}. Both use a small service handler in the trusted firmware, similar to our approach. However, they also include a relatively large operating system running in secure EL1/2. The secure operating system allows running multiple secure applications in the secure world. Since the secure world can access the normal world, the OS must trust the manufacturer's software running in secure world. Academic proposals of TEEs on the Arm architecture are plentiful. However, most proposals focus on the isolation of so-called enclaves within the secure world~\cite{winter2008trusted,santos2014using}. \cite{yaomintcb} proposes an Arm-based hardware design that separates a platform into statically-isolated domains to minimize the TCB single components have to rely on. However, this design does not provide general interrupt handling, the allocation of memory and cores is fixed at manufacturing time, and makes hardware changes that are not pragmatic for current of even future mobile phones.
The closest ARM-based design to ours is Sanctuary~\cite{brasser2019sanctuary}. Sanctuary uses similar techniques to run isolated but restricted enclaves in the normal world with the additional benefit of limiting the impact of vulnerabilities within an enclave. However, Sanctuary does not support any peripherals. Finally, recent proposals build TEEs with a trusted scheduler in TrustZone and are able to provide availability guarantees~\cite{wang2022rt}.

%% file: sections/conclusion.tex
\section{Conclusion}
Modern smartphones are controlled by various parties with different interests, including phone manufacturers, OS vendors, and users. This division of control limits users' ability to install and isolate their apps or configure peripheral access. TEEtime is a new smartphone architecture that maintains compatibility with the existing ecosystem while allowing for more balanced control among stakeholders. TEEtime makes users sovereign over their phones by enabling them to install sensitive apps in isolated domains with protected access to selected peripherals. Through this design and implementation, we demonstrated the feasibility of this approach on modern smartphone platforms.

%% file: sections/appx.tex
\section{Background on Armv8-A Architecture}
\label{appx:background}

\begin{figure}
    \centering
    \includegraphics[width=0.99\linewidth]{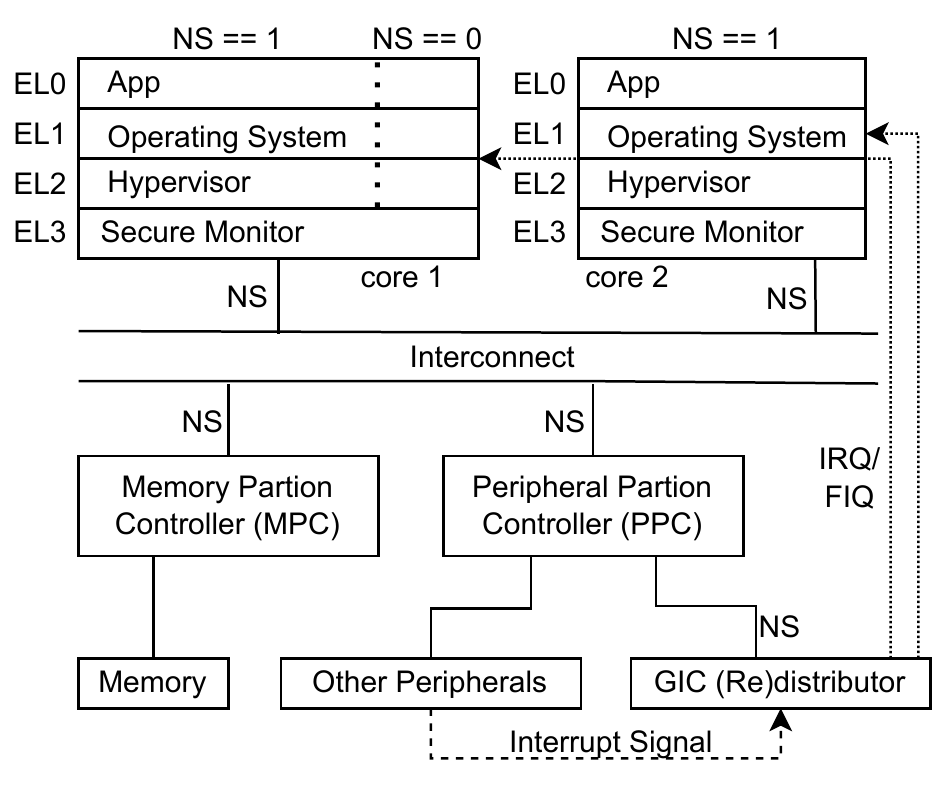}
    \caption{{\bf Conceptual Arm platform.} Core 1 is depicted with the two different states provided by the TrustZone extension, non-secure (NS == 1) and secure. This state is propagated to interconnect subordinates.}
    \label{fig:armplatform}
    \vspace{-1em}
\end{figure}

Figure \ref{fig:armplatform} depicts the a generic Arm platform with two execution cores. Armv8-A cores offer four privilege layers, from EL0 (lowest privilege) to EL3 (highest privilege). EL3 is populated by the secure monitor (SM), which is commonly considered trusted and performs low-level and highly sensitive tasks such as trusted boot, power management, and context switches between secure and non-secure state (sometimes referred to as normal state). Other privilege levels can invoke the SM and request its services through the secure monitor call (SMC) instruction. Platforms with TrustZone extensions furthermore distinguish between secure and non-secure states for EL0 to EL2. This processor state can be switched only in EL3. Software running in secure state is more privileged than and protected from software in the non-secure state. The current state of the processor is determined by the NS bit, which is propagated on buses. Commercial off-the-shelf Arm devices are commonly equipped with address space controllers (ASCs), which allow state- and sometimes core-based access control to memory (MPC) and peripherals (PPC)~\cite{cerdeira2022rezone}. MPCs can be used to divide memory address space in different regions, for which access policies can be defined separately. PPCs either work with address space partitioning of the (memory-mapped) peripherals as well, or offer per-peripheral access control. The controllers can be configured from within the secure state and EL3. An example of such controllers is the Arm TrustZone Address Space Controller (TZASC).

\section{Ideal hardware for Affinity}
\label{appx:ideal-hw}

One potential option to support the requirements from Section~\ref{sec:interrupt_iso_details} is to make hardware changes to the GIC.
It would be ideal if the GIC supported access control based on affinity: Only the core which is the target of a specific interrupt can change its routing behavior and other configuration.
This ensures that once an interrupt id is allocated to a core, only that core can configure it. 
Given such a GIC, the initial assignment can be done by EL3 software, and then at runtime, the GIC can distinguish and enforce checks.

\section{\codename on Arm Boards}
\label{appx:boards}

We procured the 4 Arm boards summarized in Table~\ref{tab:boards}. 
We had three main criteria for finding a hardware platform to prototype \codename: (a) ability to install and execute our own trusted firmware along with a custom secure monitor; 
(b) functional and documented address space controller;
(c) fine-grained configuration of the address space controller that can enforce isolation per core;
(d) address space controller that can protect peripheral and GIC memory. 

We read their documentation and performed simple tests to assess if each of them satisfies our prototype requirements. All of them have several functional peripherals. However, the existing documentation   for the address space controller and the implementation to access it from the firmware is either insufficient or completely unavailable. Without this information, we were unable to configure the protections for memory, peripherals, and the GIC. While we attempted to reverse engineer the offsets and values for configuring the controller, it quickly became untenable. This is why we decided to perform our end-to-end prototype on the Arm FVP emulator. We will continue our survey to identify a suitable board and follow-up on our requests to get the documentation from the vendors.

\begin{table*}
\centering
\caption{{\bf Summary of hardware boards.} We list the Arm cores, the GIC version, the exception levels, the size of DRAM, the security features including address space controller, peripherals, software compatibility support.}
\label{tab:boards}
\resizebox{\textwidth}{!}{%
\begin{tabular}{@{}llllll@{}}
\toprule
\textbf{Board} &
  \textbf{ARMv-A Base RevC AEM FVP} &
  \textbf{iMX8-MQ-EVK} &
  \textbf{Snapdragon HDK8450} &
  \textbf{AMD/Xilinx ZCU102} &
  \textbf{HiKey 970} \\ \midrule
CPU &
  Generic Arm-v8 and v9 Emulator &
  \begin{tabular}[c]{@{}l@{}}4x Cortex-A53 \\ 1x Cortex-M4\end{tabular} &
  \begin{tabular}[c]{@{}l@{}}SM8450P / Kryo CPU\\ Cortex-X2 / A710 / A510\end{tabular} &
  \begin{tabular}[c]{@{}l@{}}4x Cortex-A53 \\ 2x Coretx-R5F\end{tabular} &
  \begin{tabular}[c]{@{}l@{}}4x Cortex-A53 \\ 4x Cortex-A73\end{tabular} \\
GIC &
  GICv3 &
  GICv3 (GIC-500) &
  Unknown &
  GICv2 (GIC-400) &
  GICv2 (GIC-400) \\
ELs &
  EL0-EL3, all accessible &
  EL0-EL3, all accessible &
  EL3 not accessible &
  EL0-EL3, all accessible &
  EL0-EL3, all accessible \\
DRAM &
  user-defined &
  4GB DRAM &
  12GB LPDDR5 &
  4GB DRAM &
  6GB DRAM \\
Security &
  TZC-400 &
  \begin{tabular}[c]{@{}l@{}}TZC-380 (ID per full CPU)\\ Undocumented (configuration)\end{tabular} &
  \begin{tabular}[c]{@{}l@{}}Maybe TZASC? \\ Qualcomm Proprietary TEE\end{tabular} &
  \begin{tabular}[c]{@{}l@{}}HW RoT, XPPU\\ No TZASC, ID config per cluster\end{tabular} &
  \begin{tabular}[c]{@{}l@{}}TZASC(-like?) \\ Undocumented (no version, static config)\end{tabular} \\
\multirow{4}{*}{Peripherals} &
  4 UART, 8 LEDs, 8 Buttons &
  1-2 UART &
  1 UART &
  2 UART &
  7 UART, I2C, GPIO \\
 &
  Ethernet &
  Ethernet, WiFi &
  WiFi &
  Ethernet &
  Ethernet, WiFi \\
 &
  Storage &
  Bluetooth &
  Bluetooth &
  GPIO, I2C, SPI &
  GPS, Bluetooth \\
 &
  Display &
  Display (not working) &
  Touchscreen Display &
  FPGA &
  HDMI, MIPI \\
\begin{tabular}[c]{@{}l@{}}Software\\ Compatibility\end{tabular} &
  \begin{tabular}[c]{@{}l@{}}Full Root Device Access\\ Fast Deployment\end{tabular} &
  \begin{tabular}[c]{@{}l@{}}Full Root Device Access\\ Difficult deployment\end{tabular} &
  No SDK &
  Xilinx SDK both for SW and FPGA &
  Software SDK (TF+Kernel+OS) \\ \bottomrule
\end{tabular}%
}
\end{table*}
\begin{table*}
\centering
\caption{{\bf Lifecycle Breakdown.} Column 2 shows the binary sizes. Columns 3, 4, 5 show the number of cycles to launch a domain in temporal mode; Columns 6, 7, 8 show the total number of context switches, security monitor calls, and 
timer interrupts received by the domain during its boot time.
Columns 9, 10, 11 show the number of cycles to launch a domain in spatial mode.
}
\resizebox{\textwidth}{!}{%
\begin{tabular}{@{}lrrrrrrrrrr@{}}
\toprule
   & \multicolumn{1}{l}{} & \multicolumn{6}{l}{Temporal} & \multicolumn{3}{l}{Spatial} \\ \midrule
 &
  \multicolumn{1}{c}{Size} &
  \multicolumn{1}{c}{Load Binary} &
  \multicolumn{1}{c}{Setup Domain} &
  \multicolumn{1}{c}{Boot} &
  \multicolumn{1}{c}{Ctx \#} &
  \multicolumn{1}{c}{SMC \#} &
  \multicolumn{1}{c}{Timer \#} &
  \multicolumn{1}{c}{Load Binary} &
  \multicolumn{1}{c}{Setup Domain} &
  \multicolumn{1}{c}{Boot} \\ \midrule
Baremetal &   23.7 KB                   & 1,078,264    &  1,317,443  & 1,3123   &  0  & 0   & 0   &   1,063,236      &  1,324,805       &   10,538      \\
Kernel+BusyBox &   33.6 MB                   & 1,399,286,415    &  1,856,989,395  &  1,453,802,889  & 604   & 1,391,485   &  514,440  &   1,396,744,833      &    1,856,995,793     &   1,432,820,973      \\
Kernel+Buildroot &   42.0 MB                   & 1,750,611,147    &  2,321,114,735  &  1,757,813,552  & 730   &  1,683,442  &  574,144  &    1,745,354,791     &   2,321,121,981      &   1,743,338,527      \\ \bottomrule
\end{tabular}%
}
\label{tab:lifecycle}
\end{table*}

\section{Demonstration of \codename Domain Apps}
\label{appx:screenshots}

\begin{figure}
    \centering
    \includegraphics[width=\columnwidth]{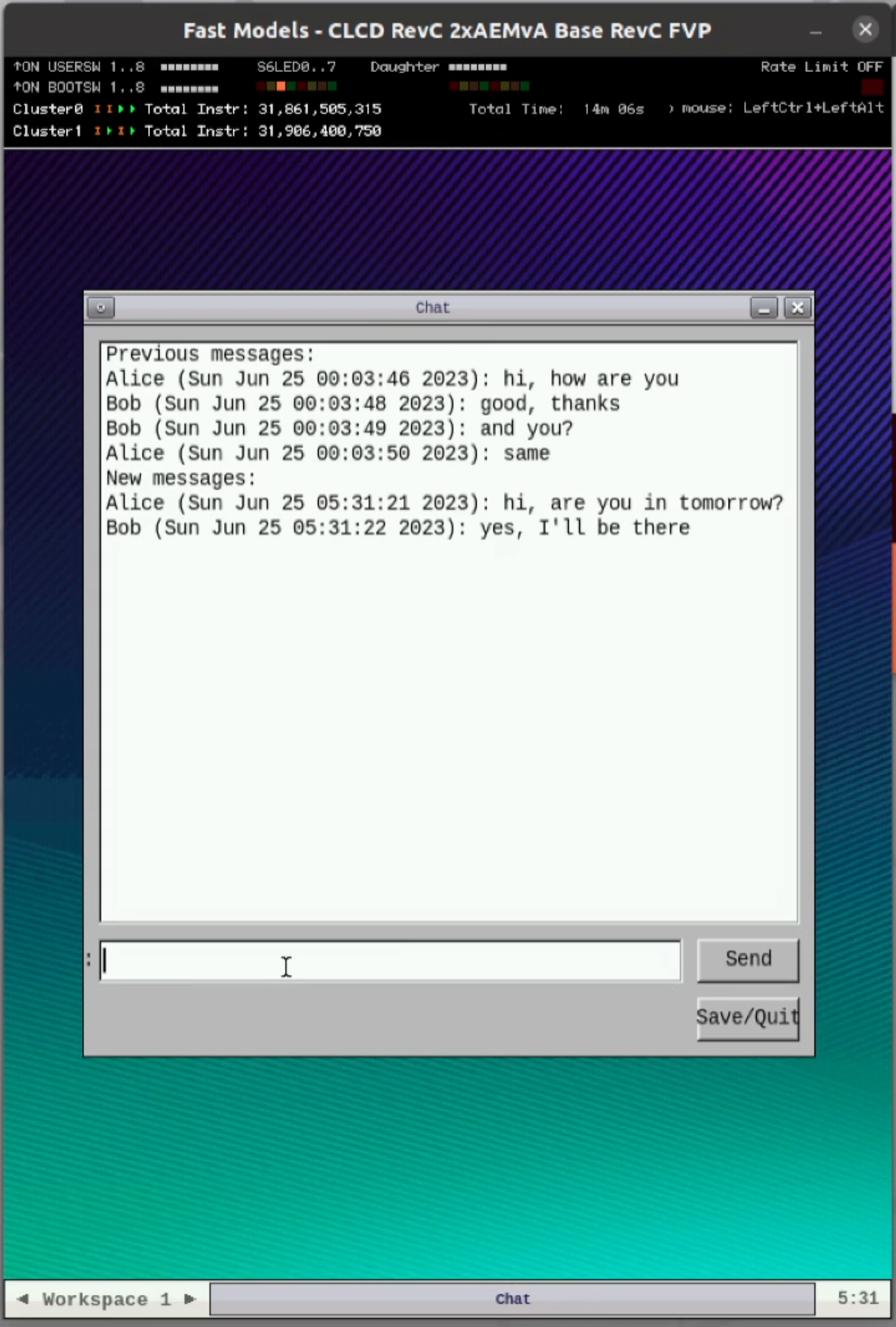}
     \caption{{\bf Secure messaging app.} The user Alice is conversing with Bob by using the mouse and keyboard to write messages and secure display to view the conversation.
     We can see Bob's responses in the app which he sent from his remote client. The previous message history was securely loaded using the storage proxy.}
    \label{fig:chat-screen}
\end{figure}

\paragraph{Secure messaging.}
Figure~\ref{fig:chat-screen} shows the screenshot of our messaging app executing in a domain.

\begin{figure}
    \centering
    \includegraphics[width=0.95\columnwidth]{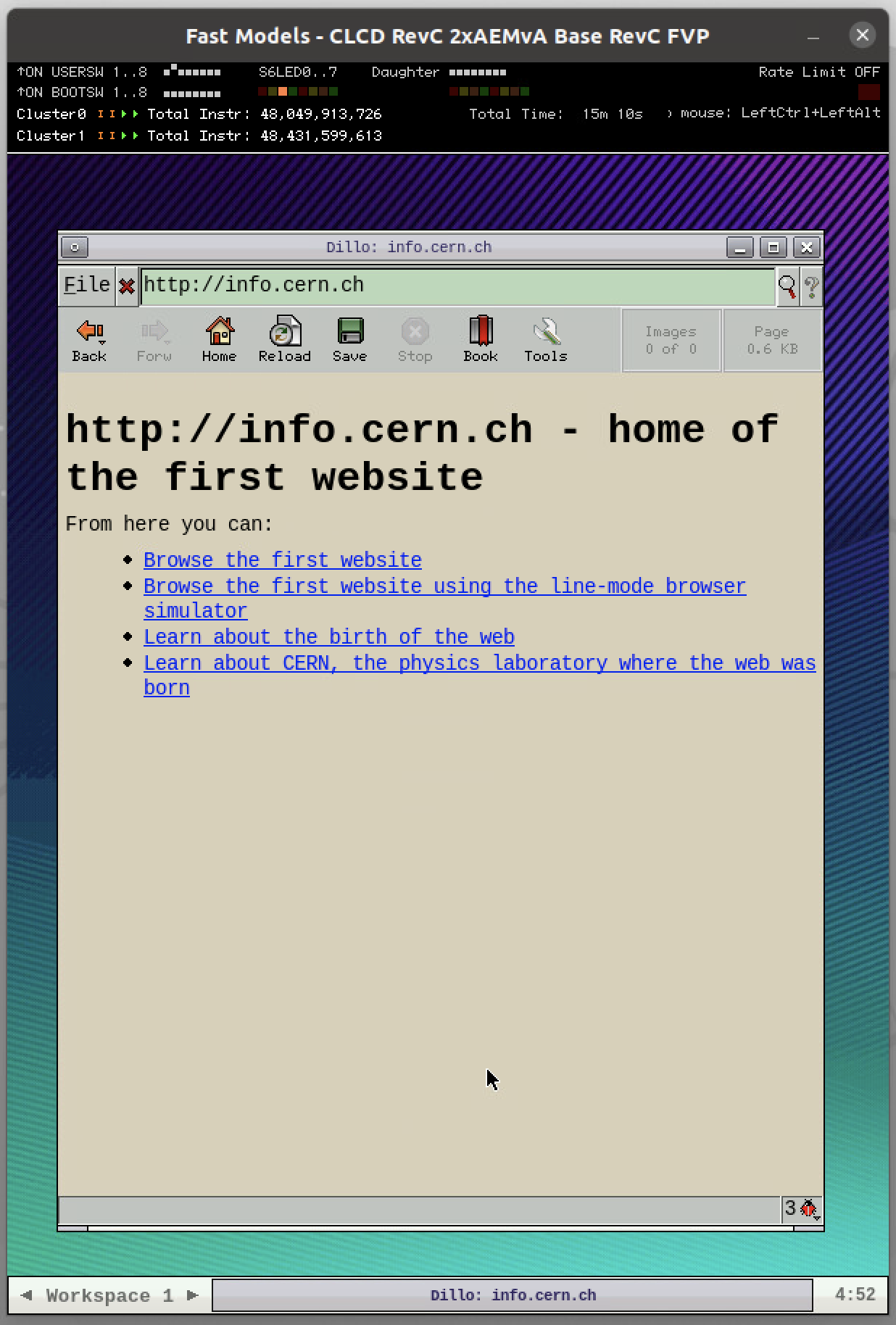}
     \caption{{\bf Secure browsing app.} User is accessing the CERN homepage to view the content on the screen.
     Network access was provided by the legacy domain via our proxylite port in the background to the app domain.}
    \label{fig:browser-screen}
\end{figure}

\paragraph{Secure Browser.}
Figure~\ref{fig:browser-screen} shows the screenshot of our web browser app executing in a domain.

\begin{figure}
    \centering
    \includegraphics[width=\columnwidth]{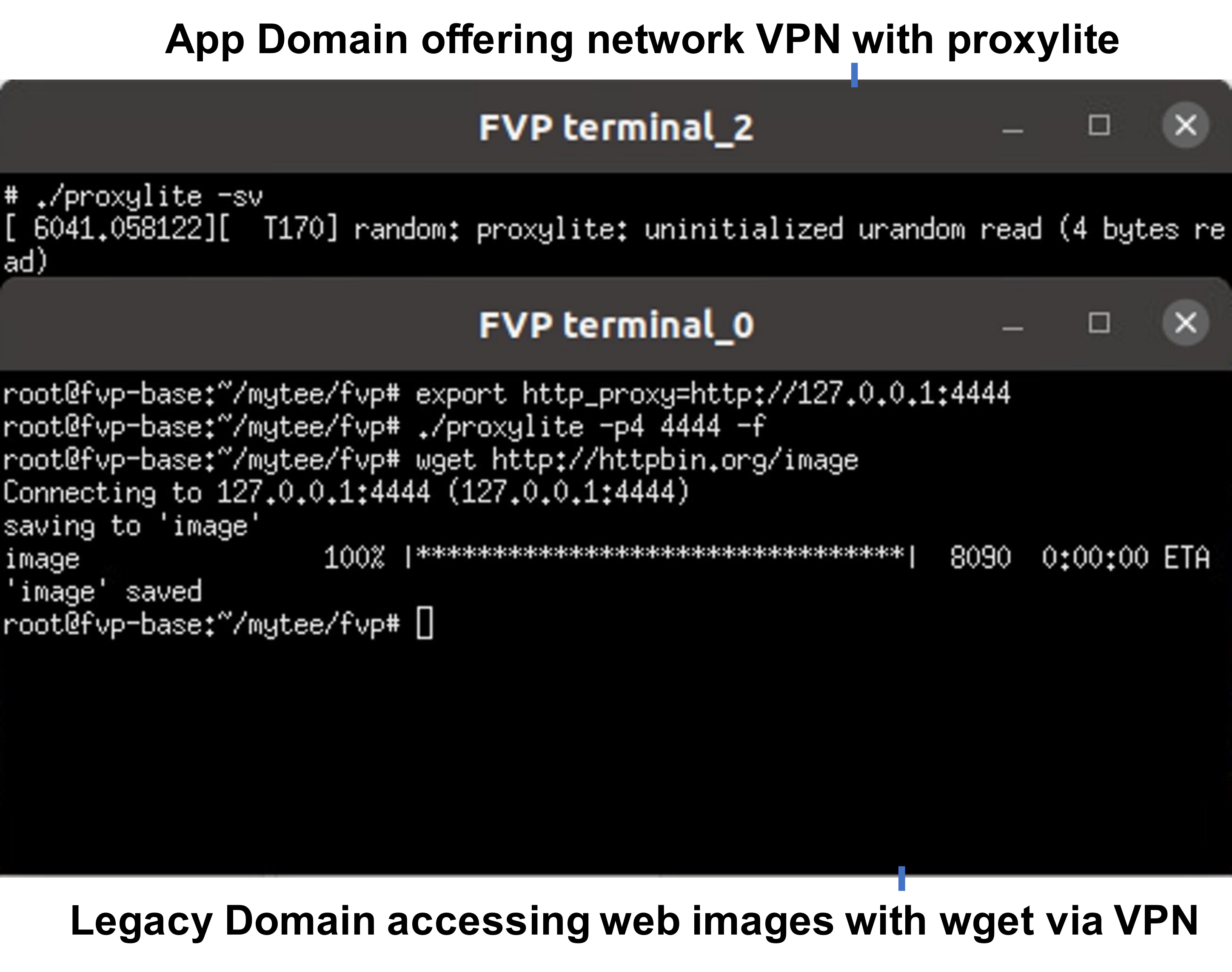}
     \caption{{\bf Secure VPN app} The user has exclusive access to the network and is offering the VPN access via proxylite to the legacy domain. 
     The legacy domain continues to perform its operations such as using $\tt{wget}$ to download images.}
    \label{fig:vpn-screen}
\end{figure}

\begin{figure}[]
    \centering
    \includegraphics[width=\columnwidth]{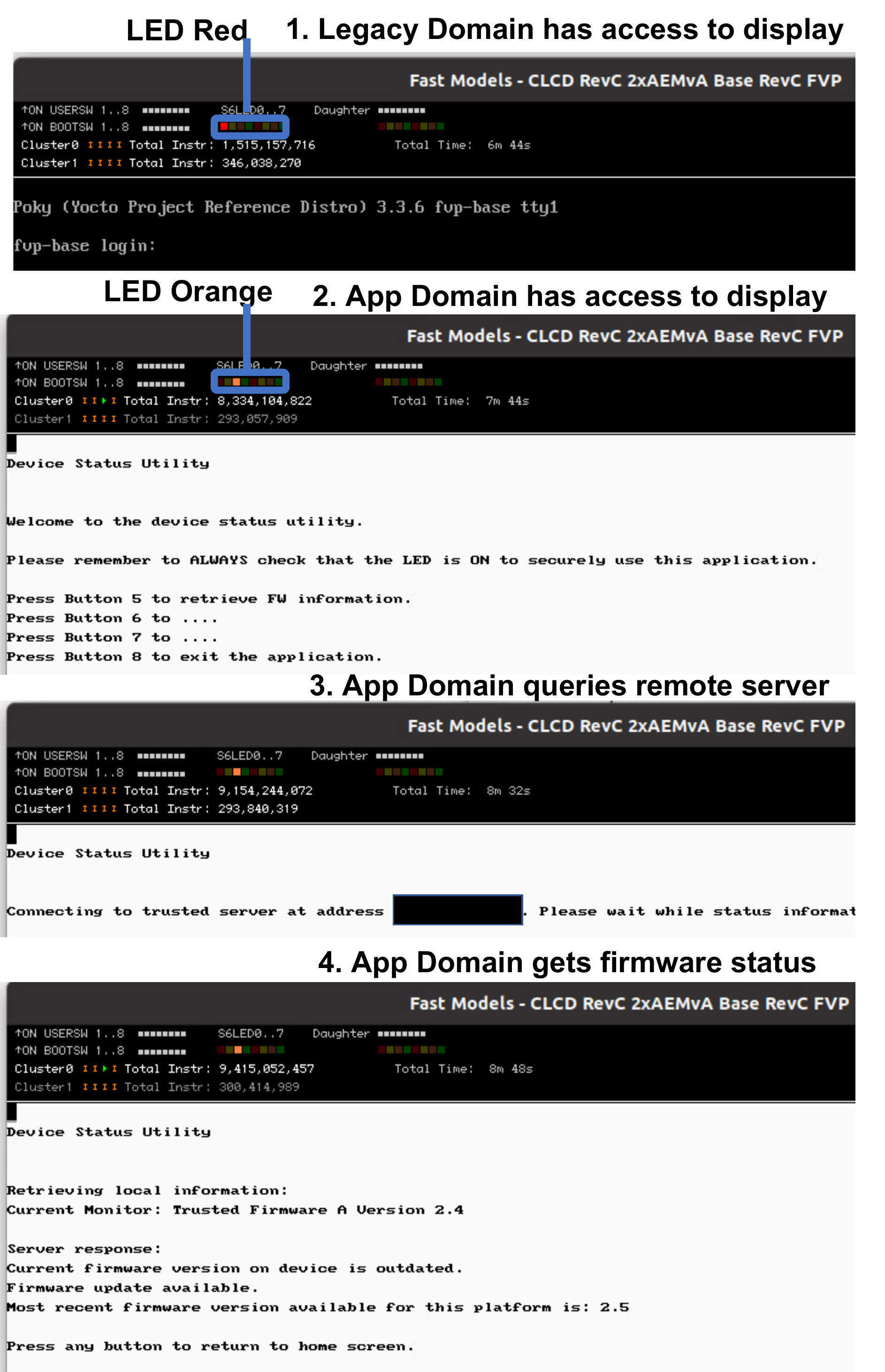}
     \caption{{\bf System Configuration \& Status app} The user has exclusive access to the buttons and display to query and view the status of firmware updates from a remote server over network with exclusive access. 
     In the first step (1) the display belongs to the legacy domain and \codename indicates this with red color LED. When the app domain is launched (2), it takes over the display  and \codename indicates this with orange color LED.
     The app domain then contacts a remote server (3) to query the configuration and firmware status (IP redacted).
     We can see the response from the server being securely displayed on the screen (4) because the app domain still has exclusive access to the display (LED still orange).
     }
    \label{fig:firmware-screen}
\end{figure}

\begin{figure}
    \centering
    \includegraphics[scale=0.27]{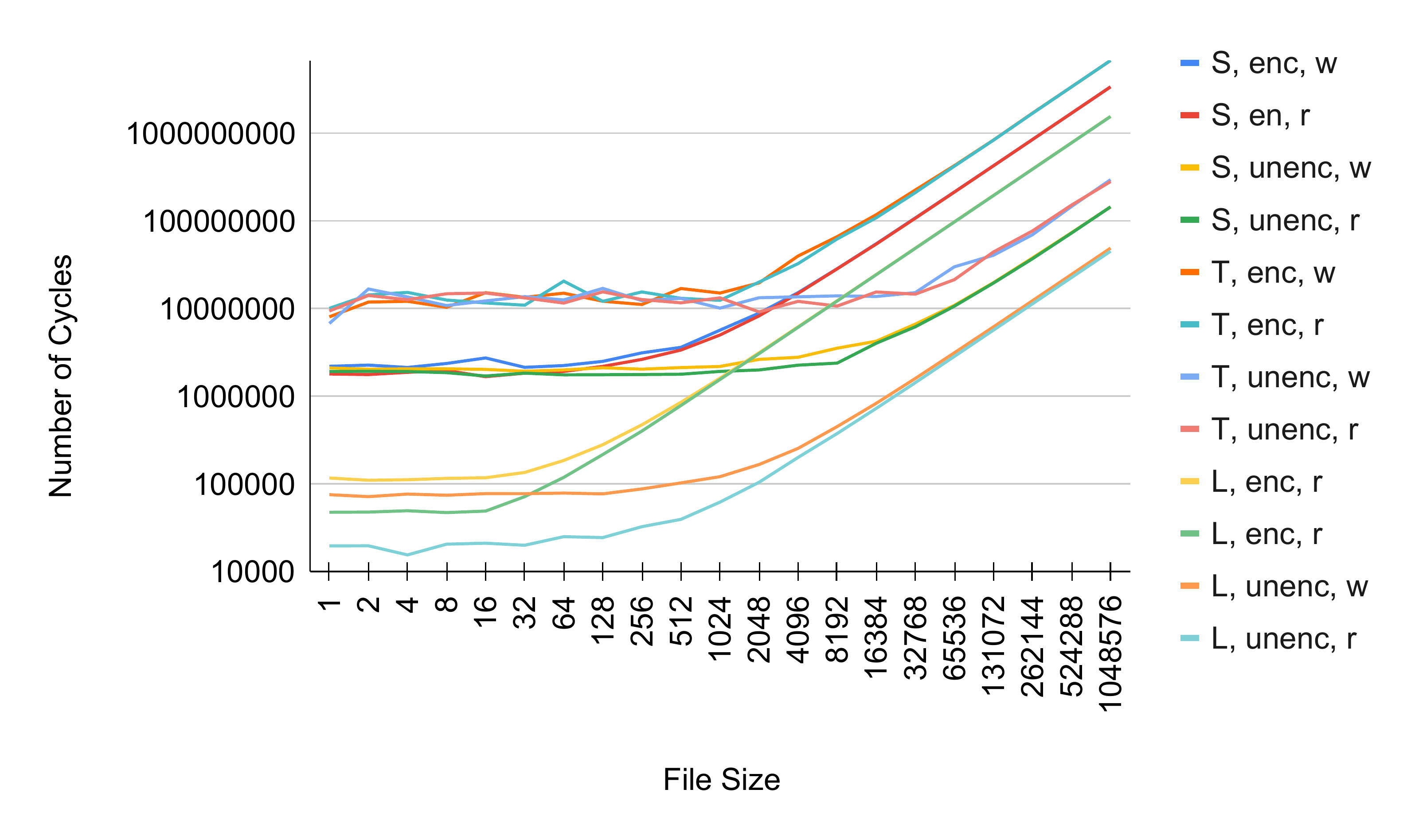}
     \caption{{\bf Storage Benchmark.} We report number of cycles to read (r) and write (w) files of different sizes. We test them in legacy (L), spatial  (S), and temporal (T) domain with encryption (enc) or without it (unenc).}
    \label{fig:storage}
\end{figure}

\begin{figure}
    \centering
    \includegraphics[scale=0.48]{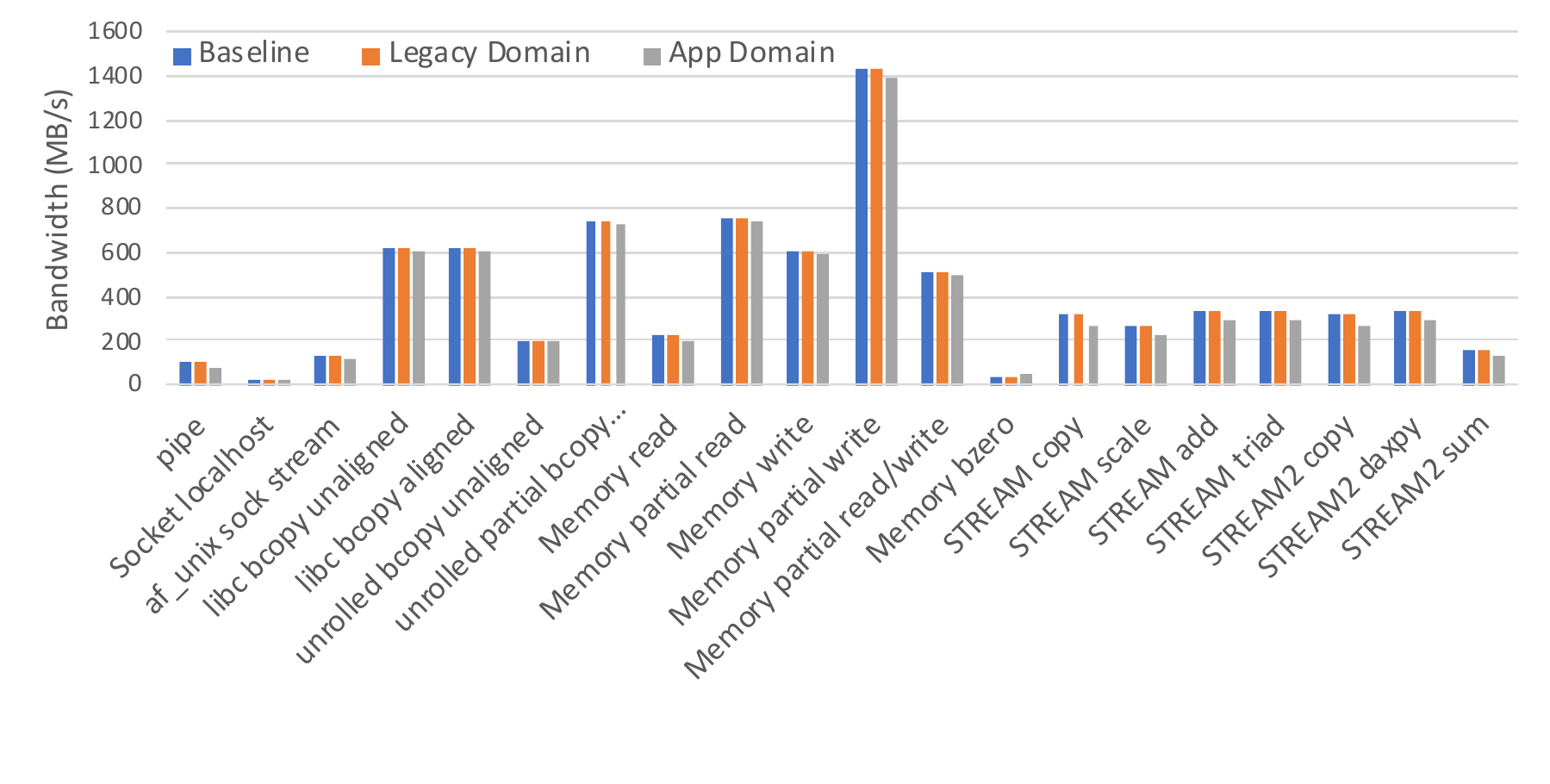}
     \caption{{\bf OS Benchmark}. We report the bandwidth of various OS operations using LMBench: performance when the legacy OS is the only domain (baseline),  performance of legacy OS when there is another  domain executing complex operations (Legacy Domain), and the performance of the app domain (App Domain).
     We exclude two tests (Memory bzero, STREAM2 fill) from the graph because they do not fit the scale.}
    \label{fig:lmbench}
\end{figure}

\paragraph{Secure VPN.}
Figure~\ref{fig:vpn-screen} shows the screenshot of our VPN app executing in a domain.

\paragraph{System Configuration \& Status.}
Figure~\ref{fig:firmware-screen} shows the screenshot of our firmware check app executing in a domain.

\section{Details of Performance Benchmarks}
\label{appx:perf}

\subsection{Storage}

Our storage benchmark performs reads and writes of different sizes through a proxy domain. We test the various sharing options, namely temporal and spatial, and also measure the baseline performance of direct storage access (L in Figure~\ref{fig:storage}). We observe that for smaller file sizes, the main contributing factor for runtime is the static overhead. For larger file sizes, this overhead amortizes, resulting in the runtime scaling linearly with the file size.

When compared to the baseline that accesses the storage in plaintext, we observe an average  slowdown of 8.60$\times$ and 3.54$\times$ for reads in temporal and spatial mode, 7.96$\times$ and 3.31$\times$ for writes. 
We attribute this mainly to memory copy operations; temporal execution is slower than spatial because of high number of context switches. 
When we turn on encryption baseline and domains, we observe an average slowdown of 2.18$\times$ for reads and writes  in spatial mode, 4.39$\times$ for reads and  writes in temporal mode.

\subsection{Network}

Table~\ref{tab:network} summarizes our results for network benchmark.
As compared to direct access, ProxyLite degrades latency, bandwidth, and throughput (B$_1$ vs B$_2$). 
Compared to ProxyLite, the domains perform better when they have exclusive network access, be it for legacy (B$_2$ vs E$_L$) or app domain (B$_2$ vs E$_A$); the app domain sees more degradation because of differences in used runtime-environments and libraries when compared to the legacy domain. %

\begin{table}
\centering
\caption{{\bf Network Benchmark.} Baseline legacy OS has to access the network directly (B$_1$) and via a proxy application running locally (B$_2$).
Exclusive network access to the legacy domain (E$_L$) and application domain (E$_A$). 
Proxy access in legacy (P$_L$) and app domain (P$_A$). 
* denotes that lower number is better. 
}
\label{tab:my-table}
\resizebox{\columnwidth}{!}{%
\begin{tabular}{@{}lrrrrrr@{}}
\toprule
\multicolumn{1}{c}{Measurement} &
  \multicolumn{1}{c}{B$_1$} &
  \multicolumn{1}{c}{B$_2$} &
  \multicolumn{1}{c}{E$_L$} &
  \multicolumn{1}{c}{E$_A$} &
  \multicolumn{1}{c}{P$_L$} &
  \multicolumn{1}{c}{P$_A$} \\ \midrule
Requests/sec         & 480.82 & 75.21   & 362.46  & 250.32  & 26.86    & 46.28  \\
Time/Request (ms)*   & 62.39 & 398.87 & 82.77  & 119.85 & 1117.06 & 648.17 \\
Transfer Rate (KB/s) & 5255.20 & 822.04  & 3961.57 & 2735.96 & 293.53   & 505.87 \\ \bottomrule
\end{tabular}%
}
\label{tab:network}
\end{table}

\subsection{System Benchmarks}

We benchmark the impact of \codename on operating system tasks. To do that, we execute a OS stress test benchmark, LMBench~\cite{lmbench}, in the legacy domain under three configurations.
We run the baseline where the OS is the only domain on the platform.
We then measure the overhead of running the OS domain 
when the app domain is executing a intensive operation (AES computation). 
Lastly, we run the benchmarks in the app domain. 
Figure~\ref{fig:lmbench} summarizes our LMBench performance.
We configure the legacy domain to have access to all peripherals, while the app domain executes on another core in spatial mode and only uses CPU and memory. 
We observe no noticable slowdown of the legacy domain (0.08\% bandwidth gain) and a minor slowdown for the app domain (9.4\% bandwidth loss).

\subsection{Lifecycle}
\label{appx:lifecycle}

Table~\ref{tab:lifecycle} summarizes the breakdown for 3 representative workloads that we execute in app domains, varying from baremetal app to full kernel boots.

%% file: main.bbl
\begin{thebibliography}{10}
\providecommand{\url}[1]{#1}
\csname url@samestyle\endcsname
\providecommand{\newblock}{\relax}
\providecommand{\bibinfo}[2]{#2}
\providecommand{\BIBentrySTDinterwordspacing}{\spaceskip=0pt\relax}
\providecommand{\BIBentryALTinterwordstretchfactor}{4}
\providecommand{\BIBentryALTinterwordspacing}{\spaceskip=\fontdimen2\font plus
\BIBentryALTinterwordstretchfactor\fontdimen3\font minus
  \fontdimen4\font\relax}
\providecommand{\BIBforeignlanguage}[2]{{%
\expandafter\ifx\csname l@#1\endcsname\relax
\typeout{** WARNING: IEEEtranS.bst: No hyphenation pattern has been}%
\typeout{** loaded for the language `#1'. Using the pattern for}%
\typeout{** the default language instead.}%
\else
\language=\csname l@#1\endcsname
\fi
#2}}
\providecommand{\BIBdecl}{\relax}
\BIBdecl

\bibitem{trusty}
{Android Open Source Project}, ``Trusty tee,''
  \url{https://source.android.com/docs/security/features/trusty}.

\bibitem{apnews_2018}
{AP News}, ``Google tracks your movements, like it or not,'' Aug 2018,
  \url{https://apnews.com/article/828aefab64d4411bac257a07c1af0ecb}.

\bibitem{apache-bench}
Apache, ``ab - apache http server benchmarking tool,''
  \url{https://httpd.apache.org/docs/2.4/programs/ab.html}, 2023.

\bibitem{appletouchid}
Apple, ``About touch id advanced security technology,''
  \url{https://support.apple.com/en-us/HT204587}.

\bibitem{applesecurity}
------, ``Apple platform security,''
  \url{https://support.apple.com/guide/security/welcome/web}.

\bibitem{applesecureenclave}
------, ``Secure enclave,''
  \url{https://support.apple.com/guide/security/secure-enclave-sec59b0b31ff/web}.

\bibitem{arm_cca}
Arm, ``Arm architecture reference manual supplement, the realm management
  extension (rme), for armv9-a,''
  \url{https://developer.arm.com/documentation/ddi0615}.

\bibitem{arm_gicv3}
{Arm}, ``Arm generic interrupt controller architecture specification,''
  \url{https://developer.arm.com/documentation/ihi0069/h/}.

\bibitem{armfvp}
Arm, ``Fixed virtual platforms,'' \url{https://developer.arm.com/Tools and
  Software/Fixed virtual Platforms}.

\bibitem{armtzc400}
------, ``{ARM CoreLink TZC-400 TrustZone} address space controller technical
  reference manual,'' ARM, Tech. Rep., 2014.

\bibitem{arstechnica_2018}
{Ars Technica}, ``Googles iron grip on {A}ndroid: Controlling open source by
  any means necessary,'' Jul 2018,
  \url{https://arstechnica.com/gadgets/2018/07/googles-iron-grip-on-android-controlling-open-source-by-any-means-necessary/}.

\bibitem{bbcnews_2019}
{BBC News}, ``Apple bans {H}ong {K}ong protest location app,'' Oct 2019,
  \url{https://www.bbc.com/news/technology-49919459}.

\bibitem{bock2019geneva}
K.~Bock, G.~Hughey, X.~Qiang, and D.~Levin, ``Geneva: Evolving censorship
  evasion strategies,'' in \emph{Proceedings of the 2019 ACM SIGSAC Conference
  on Computer and Communications Security}, 2019, pp. 2199--2214.

\bibitem{brasser2019sanctuary}
F.~Brasser, D.~Gens, P.~Jauernig, A.-R. Sadeghi, and E.~Stapf, ``Sanctuary:
  {ARM}ing {T}rust{Z}one with user-space enclaves.'' in \emph{NDSS}, 2019.

\bibitem{buildroot}
{Buildroot Association}, ``{Buildroot},'' \url{https://buildroot.org/}.

\bibitem{busch2020unearthing}
M.~Busch, J.~Westphal, and T.~M{\"u}ller, ``Unearthing the trustedcore: A
  critical review on huawei's trusted execution environment.'' in \emph{WOOT@
  USENIX Security Symposium}, 2020.

\bibitem{businessinsider_2019}
{Business Insider}, ``A security expert found that {A}pple's latest i{P}hone
  can still track your location data, even if you toggle it off for every
  app,'' Dec 2019,
  \url{https://www.businessinsider.com/apple-iphone-11-pro-collects-location-data-krebs-report-2019-12?r=US&IR=T}.

\bibitem{kai2019guard}
L.~Cai, ``Guard your data with the qualcomm snapdragon mobile platform,''
  Qualcomm, Tech. Rep., 2019.

\bibitem{gunyah_qualcomm}
Q.~I. Center, ``{Gunyah Hypervisor},''
  \url{https://github.com/quic/gunyah-hypervisor}.

\bibitem{cerdeira2022rezone}
D.~Cerdeira, J.~Martins, N.~Santos, and S.~Pinto, ``Rezone: Disarming trustzone
  with tee privilege reduction,'' \emph{arXiv preprint arXiv:2203.01025}, 2022.

\bibitem{dillo}
J.~A. Cid, ``{Dillo},'' \url{https://www.dillo.org/}.

\bibitem{delugre2011reverse}
G.~Delugr{\'e}, ``Reverse engineering a qualcomm baseband,'' in \emph{28th
  Chaos Communication Congress}, 2011.

\bibitem{deng2022strongbox}
Y.~Deng, C.~Wang, S.~Yu, S.~Liu, Z.~Ning, K.~Leach, J.~Li, S.~Yan, Z.~He,
  J.~Cao \emph{et~al.}, ``Strongbox: A gpu tee on arm endpoints,'' in
  \emph{Proceedings of the 2022 ACM SIGSAC Conference on Computer and
  Communications Security}, 2022, pp. 769--783.

\bibitem{fido_spec}
{FIDO Alliance}, ``Client to authenticator protocol,''
  \url{https://fidoalliance.org/specs/fido-v2.1-ps-20210615/fido-client-to-authenticator-protocol-v2.1-ps-errata-20220621.html}.

\bibitem{forbes_2021}
Forbes, ``Parler at risk of going offline after bans from {A}mazon, {A}pple and
  {G}oogle,'' Jan 2021,
  \url{https://www.forbes.com/sites/jemimamcevoy/2021/01/10/parler-at-risk-of-going-offline-after-bans-from-amazon-apple-and-google/}.

\bibitem{androidsecuritybulletin}
Google, ``Android security bulletins,''
  \url{https://source.android.com/docs/security/bulletin}.

\bibitem{pkvm}
------, ``Avf architecture,''
  \url{https://source.android.com/docs/core/virtualization/architecture}.

\bibitem{google_gms}
------, ``Google mobile services,'' \url{https://www.android.com/gms/}.

\bibitem{hafnium_hypervisor}
------, ``{Hafnium architecture},''
  \url{https://hafnium.googlesource.com/hafnium/+/HEAD/docs/Architecture.md}.

\bibitem{google_passkeys}
------, ``Passkeys use cases,''
  \url{https://developers.google.com/identity/passkeys/use-cases}.

\bibitem{androidsecurity}
------, ``Secure android devices,''
  \url{https://source.android.com/docs/security/}.

\bibitem{proxy-lite}
B.~Jesus, ``{ProxyLite},'' \url{https://github.com/00cpxxx/proxylite}.

\bibitem{fluxbox}
H.~Kinnunen, ``{Fluxbox},'' \url{http://fluxbox.org/}.

\bibitem{larus2021whose}
J.~R. Larus, ``Whose smartphone is it?'' \emph{Communications of the ACM},
  vol.~64, no.~9, pp. 41--42, 2021.

\bibitem{lmbench}
L.~McVoy and C.~Staelin, ``Lmbench: Portable tools for performance analysis,''
  in \emph{ATC}.\hskip 1em plus 0.5em minus 0.4em\relax USA: USENIX
  Association, 1996.

\bibitem{grapheneos}
D.~Micay, ``{GrapheneOS},'' \url{https://grapheneos.org/}.

\bibitem{imx8m}
{NXP}, ``{i.MX 8M Dual/8M QuadLite/8M Quad Applications Processors Reference
  Manual},'' NXP, Tech. Rep., 2023.

\bibitem{rane2015raccoon}
A.~Rane, C.~Lin, and M.~Tiwari, ``Raccoon: Closing digital side-channels
  through obfuscated execution,'' in \emph{24th $\{$USENIX$\}$ Security
  Symposium ($\{$USENIX$\}$ Security 15)}, 2015, pp. 431--446.

\bibitem{reuters_2017}
Reuters, ``{G}oogle, {A}pple remove {N}avalny app from stores as russian
  elections begin,'' Sep 2017,
  \url{https://www.reuters.com/world/europe/google-apple-remove-navalny-app-stores-russian-elections-begin-2021-09-17/}.

\bibitem{reuters_2020_b}
------, ``Denmark angry at {G}oogle censorship of some {D}anish content, seeks
  talks,'' Aug 2020,
  \url{https://www.re}uters.com/article/us-google-censorship-denmark-idUSKCN2561TG.

\bibitem{reuters_2020}
------, ``Germany at odds with {A}pple on smartphone coronavirus contact
  tracing,'' Apr 2020,
  \url{https://www.reuters.com/article/us-health-coronavirus-europe-tech/germany-at-odds-with-apple-on-smartphone-coronavirus-contact-tracing-idUSKCN2251MR}.

\bibitem{samsung-pay}
Samsung, ``Samsung pay,'' \url{https://www.samsung.com/us/samsung-pay/}.

\bibitem{santos2014using}
N.~Santos, H.~Raj, S.~Saroiu, and A.~Wolman, ``Using arm trustzone to build a
  trusted language runtime for mobile applications,'' in \emph{Proceedings of
  the 19th international conference on Architectural support for programming
  languages and operating systems}, 2014, pp. 67--80.

\bibitem{schechter2007emperor}
S.~E. Schechter, R.~Dhamija, A.~Ozment, and I.~Fischer, ``The emperor's new
  security indicators,'' in \emph{2007 IEEE Symposium on Security and Privacy
  (SP'07)}.\hskip 1em plus 0.5em minus 0.4em\relax IEEE, 2007, pp. 51--65.

\bibitem{schmidt2018google}
D.~Schmidt, ``Google data collection,'' Vanderbilt University, Tech. Rep.,
  2018,
  \url{http://www.dre.vanderbilt.edu/~schmidt/PDF/google-data-collection.pdf}.

\bibitem{sierratee}
Sierraware, ``Sierratee trusted execution environment,''
  \url{https://www.sierraware.com/open-source-ARM-TrustZone.html}.

\bibitem{soriente2007beda}
C.~Soriente, G.~Tsudik, and E.~Uzun, ``Beda: Button-enabled device pairing,''
  \emph{Cryptology ePrint Archive}, 2007.

\bibitem{fltk}
B.~Spitzak, ``{Fast Light Toolkit},'' \url{https://www.fltk.org/}.

\bibitem{sun2015trustice}
H.~Sun, K.~Sun, Y.~Wang, J.~Jing, and H.~Wang, ``Trustice: Hardware-assisted
  isolated computing environments on mobile devices,'' in \emph{2015 45th
  Annual IEEE/IFIP International Conference on Dependable Systems and
  Networks}.\hskip 1em plus 0.5em minus 0.4em\relax IEEE, 2015, pp. 367--378.

\bibitem{telegraph_2018}
T.~Telegraph, ``Fortnite decision to bypass the {G}oogle {P}lay store sparks
  security concerns for teen gamers,'' Aug 2018,
  \url{https://www.telegraph.co.uk/technology/2018/08/08/fortnite-decision-bypass-google-play-store-sparks-security-concerns/}.

\bibitem{theguardian_2020_a}
{The Guardian}, ``France urges {A}pple and {G}oogle to ease privacy rules on
  contact tracing,'' Apr 2020,
  \url{https://www.theguardian.com/world/2020/apr/21/france-apple-google-privacy-contact-tracing-coronavirus}.

\bibitem{theguardian_2020_b}
------, ``{NHS} in standoff with {A}pple and {G}oogle over coronavirus
  tracing,'' Apr 2020,
  \url{https://www.theguardian.com/technology/2020/apr/16/nhs-in-standoff-with-apple-and-google-over-coronavirus-tracing}.

\bibitem{thenewyorktimes_2020}
{The New York Times}, ``Fortnite creator sues {A}pple and {G}oogle after ban
  from app stores,'' Aug 2020,
  \url{https://www.nytimes.com/2020/08/13/technology/apple-fortnite-ban.html}.

\bibitem{theverge_2016}
{The Verge}, ``Google removes {S}amsung's first {A}ndroid ad blocker from the
  {P}lay {S}tore,'' Feb 2016,
  \url{https://www.theverge.com/2016/2/3/10905672/google-samsung-adblock-fast-android-ad-blocker-removal}.

\bibitem{theverge_2020}
------, ``Apple just kicked {F}ortnite off the app store,'' Aug 2020,
  \url{https://www.theverge.com/2020/8/13/21366438/apple-fortnite-ios-app-store-violations-epic-payments}.

\bibitem{optee}
{TrustedFirmware Project}, ``Op-tee,'' \url{https://www.op-tee.org/}.

\bibitem{trustonic}
Trustonic, ``Trustonic tee,'' \url{https://www.trustonic.com/technology/}.

\bibitem{busybox}
D.~Vlasenko, ``{Busybox},'' \url{https://busybox.net/}.

\bibitem{gdpr}
P.~Voigt and A.~Von~dem Bussche, ``The {EU} general data protection regulation
  ({GDPR}),'' \emph{A Practical Guide, 1st Ed., Cham: Springer International
  Publishing}, vol.~10, no. 3152676, pp. 10--5555, 2017.

\bibitem{wang2022rt}
J.~Wang, A.~Li, H.~Li, C.~Lu, and N.~Zhang, ``Rt-tee: Real-time system
  availability for cyber-physical systems using arm trustzone,'' in \emph{2022
  IEEE Symposium on Security and Privacy (SP)}.\hskip 1em plus 0.5em minus
  0.4em\relax IEEE, 2022, pp. 352--369.

\bibitem{wang2020symtcp}
Z.~Wang and S.~Zhu, ``Symtcp: eluding stateful deep packet inspection with
  automated discrepancy discovery,'' in \emph{Network and Distributed System
  Security Symposium (NDSS)}, 2020.

\bibitem{winter2008trusted}
J.~Winter, ``Trusted computing building blocks for embedded linux-based arm
  trustzone platforms,'' in \emph{Proceedings of the 3rd ACM workshop on
  Scalable trusted computing}, 2008, pp. 21--30.

\bibitem{xilinx_ultrascale}
{Xilinx}, ``{Zynq UltraScale+ Device Technical Reference Manual},'' Xilinx,
  Tech. Rep., 2023.

\bibitem{xorg}
{X.Org Foundation}, ``{Xorg},'' \url{https://www.x.org/}.

\bibitem{yaomintcb}
\BIBentryALTinterwordspacing
Z.~Yao, S.~M. Seyed~Talebi, M.~Chen, A.~Amiri~Sani, and T.~Anderson,
  ``Minimizing a smartphone's tcb for security-critical programs with
  exclusively-used, physically-isolated, statically-partitioned hardware,'' in
  \emph{Proceedings of the 21st Annual International Conference on Mobile
  Systems, Applications and Services}, ser. MobiSys '23.\hskip 1em plus 0.5em
  minus 0.4em\relax New York, NY, USA: Association for Computing Machinery,
  2023, p. 233–246. [Online]. Available:
  \url{https://doi.org/10.1145/3581791.3596864}
\BIBentrySTDinterwordspacing

\bibitem{yu2022elasticlave}
J.~Z. Yu, S.~Shinde, T.~E. Carlson, and P.~Saxena, ``Elasticlave: An efficient
  memory model for enclaves,'' in \emph{31st USENIX Security Symposium (USENIX
  Security 22)}, 2022, pp. 4111--4128.

\bibitem{attackingandroidfromtz}
T.~Zahavi-Brunner, ``Attacking the android kernel using the qualcomm
  trustzone,''
  \url{https://tamirzb.com/attacking-android-kernel-using-qualcomm-trustzone}.

\bibitem{zhang2016truspy}
N.~Zhang, K.~Sun, D.~Shands, W.~Lou, and Y.~T. Hou, ``Truspy: Cache
  side-channel information leakage from the secure world on arm devices,''
  \emph{Cryptology ePrint Archive}, 2016.

\end{thebibliography}
